%% file: main.tex
\documentclass[12pt,twoside]{mitthesis}
\usepackage{lgrind}
\usepackage{amsmath} 
\usepackage{amssymb} 
\usepackage{tcolorbox} 
\usepackage{tabularx} 
\usepackage{colortbl}  
\tcbuselibrary{skins}  

\usepackage{xcolor} 
\usepackage{color, colortbl} 
\definecolor{Gray}{gray}{0.9}
\usepackage{caption}
\usepackage{multirow,array} 

\tcbset{tab1/.style={fonttitle=\bfseries\large,fontupper=\normalsize\sffamily,
colback=yellow!10!white,colframe=red!75!black,colbacktitle=white!40!white,
coltitle=black,center title,freelance,frame code={
\foreach \n in {north east,north west,south east,south west}
{\path [fill=white!75!black] (interior.\n) circle (3mm); };},}}

\tcbset{tab2/.style={enhanced,fonttitle=\bfseries,fontupper=\normalsize,
colback=white!10!white,colframe=white!0!white,colbacktitle=white!40!white,
coltitle=black,center title}}

\usepackage [utf8] {inputenc} 
\usepackage{graphicx} 
\usepackage{epstopdf}  
\usepackage{cmap}
\usepackage[T1]{fontenc}
\usepackage{mathptmx}  
\usepackage{titlesec}
\titleformat{\chapter}[display]   
  {\centering\normalfont\huge\bfseries}
  {\chaptertitlename\ \thechapter}{17pt}{\Huge}
\usepackage{lipsum}
\usepackage{fancyhdr}
\usepackage{float} 

\ifx\files\all  \else   \fi

\begin{document}
\setcounter{secnumdepth}{4}
\pagestyle{empty}
\include{cover1}  
\pagestyle{plain}
\pagenumbering{roman}
\clearpage
\setcounter{page}{2}
\include{cover2}
\include{cover3}
\include{cover7}
\include{cover9}
\let\cleardoublepage\clearpage

\include{contents}

\include{cover10}
\include{cover11}
\pagenumbering{arabic}
\pagestyle{fancy}
\fancyhf{}
\renewcommand{\chaptermark}[1]{\markboth{\thechapter. #1}{}}
\renewcommand{\sectionmark}[1]{\markright{\thesection. #1}}
\renewcommand{\subsectionmark}[1]{\markright{\thesubsection. #1}}
\lhead[\fancyplain{}{}]%
      {\fancyplain{}{\bfseries\rightmark}}
\rhead{\bfseries Chapter \thechapter}
\cfoot{\bfseries\thepage}
\tcbset{tab1/.style={fonttitle=\bfseries\large,fontupper=\normalsize\sffamily,
colback=yellow!10!white,colframe=red!75!black,colbacktitle=white!40!white,
coltitle=black,center title,freelance,frame code={
\foreach \n in {north east,north west,south east,south west}
{\path [fill=white!75!black] (interior.\n) circle (3mm); };},}}

\tcbset{tab2/.style={enhanced,fonttitle=\bfseries,fontupper=\normalsize,
colback=white!10!white,colframe=black!50!black,colbacktitle=white!40!white,
coltitle=black,center title}}

\include{chap1}
\let\cleardoublepage\clearpage
\include{chap2}
\let\cleardoublepage\clearpage
\newcolumntype{Y}{>{\centering\arraybackslash}X}
\include{chap3}
\let\cleardoublepage\clearpage
\include{chap4}
\let\cleardoublepage\clearpage
\include{chap5}
\let\cleardoublepage\clearpage
\pagestyle{plain}
\include{biblio}
\end{document}

%% file: cover1.tex

\begin{center}

\LARGE Considering Fading Effects for Vertical Handover in Heterogenous Wireless Networks

\end{center}

\begin{figure}[h]
  \centering
  \includegraphics[scale=0.3]{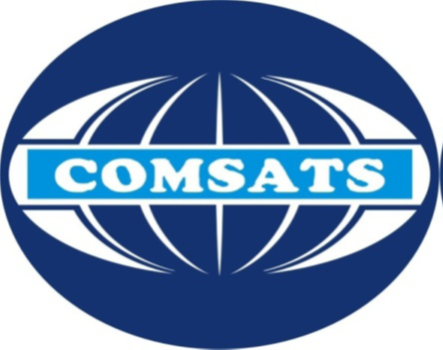}\\

\end{figure}

\begin{center}

\emph{\large By}\\

\Large Omoniwa Babatunji\\
\Large CIIT/FA13-RCE-001/ISB\footnote{© Copyright By Omoniwa Babatunji. All Rights Reserved}\\
\vfill
\Large MS Thesis\\
\Large In\\
\Large Computer Engineering
\end{center}
\vfill

\begin{center}
\Large COMSATS Institute of Information Technology\\
\Large Islamabad – Pakistan
\end{center}

\begin{center}
\LARGE Fall, 2014
\end{center}

%% file: cover2.tex

\begin{minipage}{0.1\textwidth}
  \includegraphics[scale=0.2]{ciitlogo.png}
\end{minipage}
\begin{minipage}{0.9\textwidth}\raggedleft
\textbf{\Large COMSATS Institute of Information Technology}
\end{minipage}
\vfill
\vfill
\begin{center}

\LARGE Considering Fading Effects for Vertical Handover in Heterogenous Wireless Networks
\end{center}

\vfill
\begin{center}\large
A Thesis Presented to\\
\vfill
\Large COMSATS Institute of Information Technology, Islamabad
\end{center}

\vfill
\begin{center}\normalsize
In partial fulfillment\\
of the requirement for the degree of
\end{center}

\begin{center}\LARGE
MS (Computer Engineering)
\end{center}

\vfill
\begin{center}

\normalsize © Copyright By\\

\large Omoniwa Babatunji\\
\large CIIT/FA13-RCE-001/ISB\\

All Rights Reserved
\end{center}
\vfill

\begin{center}
\Large Fall, 2014
\end{center}

%% file: cover3.tex

\begin{center}

\LARGE Considering Fading Effects for Vertical Handover in Heterogenous Wireless Networks
\end{center}
\noindent\rule{15.5cm}{3pt}\\

\normalsize
A Post Graduate Thesis submitted to the Department of Electrical Engineering as partial fulfillment of the requirement for the award of Degree of MS (Computer Engineering).\\
\vfill
\begin{center}
\begin{table}[ht]
\centering

\begin{tabular}{!{\vrule width 2pt} >{\centering}p{2.3in}!{\vrule width 2pt}c!{\vrule width 2pt}}
\hline \noalign{\hrule height 2pt}
{\fontsize{16}{15} \normalfont\cellcolor{black!50} } & {\fontsize{16}{15} \normalfont\cellcolor{black!50} } \\
{\fontsize{16}{15} \normalfont \cellcolor{black!50}Name} & {\fontsize{16}{15} \normalfont \cellcolor{black!50} Registration Number} \\ \hline
& \\
{\fontsize{16}{15} \normalfont Omoniwa Babatunji} & {\fontsize{16}{15} \normalfont CIIT/FA13-RCE-001/ISB}\\[2ex] \noalign{\hrule height 2pt}
\end{tabular}
\end{table}
\end{center}
\vfill
\noindent \textbf{\large Supervisor:}
\vfill
\normalsize
\noindent Dr. Riaz Hussain\\
Assistant Professor,\\
Department of Electrical Engineering,\\
COMSATS Institute of Information Technology (CIIT),\\
Islamabad Campus.\\
December, 2014.

%% file: cover7.tex

\addcontentsline{toc}{chapter}{Dedication}
\vspace{0.7cm}
\begin{center}
{\fontsize{16}{15} \bf DEDICATION}\\
\end{center}

\vspace{1cm}
\begin{center}
  \large{ { {\Huge $\mathcal{D}$}edicated

      to my lovely wife, Janet, \\
      who sacrificed her time and career\\
      to accompany me for this study.
      } }
\end{center}
\clearpage
\newpage

%% file: cover9.tex

\addcontentsline{toc}{chapter}{Abstract} 
\begin{center}
{{\fontsize{16}{15} \bf ABSTRACT \\ \vspace{0.3cm} \normalfont Considering Fading Effects for Vertical Handover in Heterogenous Wireless Networks} \\}
\vspace{0.4cm}
\end{center}
\normalsize
Over the years, vertical handover has attracted the interest of numerous researchers. Despite the attractive benefits of integrating different wireless platforms, mobile users are confronted with the issue of detrimental handover. As a mobile node (MN) moves within a heterogeneous environment, satisfactory quality of service (QoS) is desired by ensuring efficient vertical handover. This demands not only the efficient execution of vertical handover, but also optimized pre-handover decisions, such as: handover necessity estimation (HNE), handover triggering condition estimation (HTCE) and handover target selection (HTS). The existing works on HNE and HTCE optimization considered the coverage region of a point of attachment to be circular, ignoring the fading effect. This paper considers the effect of shadow fading and used extensive geometric and probability analysis in modelling the coverage area of a WLAN cell. Thus, presents a realistic and novel model with an attempt to ensure optimal handover as a mobile node (MN) traverses a heterogeneous wireless environment.

In the proposed HNE approach, the dwell time is estimated along with the threshold values to ensure an optimal handover decision by the MN, while the probability of unnecessary handover and handover failure are kept within tolerable bounds. The proposed HTCE approach estimates the optimal handover triggering point at which an MN will need to initiate a handover in order to avoid connection breakdowns as well as maximize the usage of the preferred access network. In the proposed HTS approach, Grey Relational Analysis (GRA) algorithm was applied on two different case studies and used to select an optimal access network to perform a handover based on certain performance criteria. Monte-Carlo simulations were carried out to show the behaviour of the proposed HNE and HTCE models. Results were validated by comparing the proposed models with existing works.

\clearpage
\newpage

%% file: contents.tex

\setcounter{secnumdepth}{4}
\tableofcontents
\let\cleardoublepage\clearpage
\newpage
\listoffigures
\let\cleardoublepage\clearpage
\newpage
\listoftables
\let\cleardoublepage\clearpage

%% file: cover10.tex

\vspace{-1cm}
\addcontentsline{toc}{chapter}{List of Symbols}
\chapter*{List of Symbols}
\markright{List of Symbols}
\markright{List of Symbols}
\vspace{-1cm}
\newcolumntype{Y}[1]{>{\raggedright\arraybackslash}p{#1}}
\begin{table}[H]
\begin{tcolorbox}[tab2,tabularx={X||p{9cm}}]

\begin{flushleft}
\textbf{\Large{\textit{Symbol}}}
\end{flushleft}

 &
\begin{flushleft}
\textbf{\Large{}}
\end{flushleft}

      \\\hline\hline
$\tau$ & Handover latency  \\\hline
$\tau_A$  & Handover latency for moving-in  \\\hline
$\tau_D$  & Handover latency for moving-out \\\hline
$\tau_T$  & Total handover latency \\\hline
$\mu$  & Mean \\\hline
$\sigma$  & Standard deviation \\\hline
$\theta$  & Traversal angle of the MN \\\hline
$P_f$  & Probability of handover failure \\\hline
$P_u$  & Probability of unnecessary handover \\\hline
$P_{Break}$  & Probability of connection breakdown \\\hline
$\beta$  & Path loss exponent \\\hline
$\zeta$  & Distinguishing coefficient \\\hline
$\gamma$  & Grey relational coefficient \\\hline
$\Gamma$  & Grey relational grade \\

\end{tcolorbox}
\end{table}
%
%
%
%
%
%
%
%
%
%
%
%

\clearpage

%% file: cover11.tex

\vspace{-1cm}
\addcontentsline{toc}{chapter}{List of Acronyms}
\chapter*{List of Acronyms}
\markright{List of Symbols}
\vspace{-1cm}

\newcolumntype{Y}[1]{>{\raggedright\arraybackslash}p{#1}}
\begin{table}[H]
\begin{tcolorbox}[tab2,tabularx={X||p{9cm}}]

\begin{flushleft}
\textbf{\Large{\textit{Acronym}}}
\end{flushleft}

 &
\begin{flushleft}
\textbf{\Large{}}
\end{flushleft}

      \\\hline\hline
1G,2G,3G,4G & Generation of mobile communication   \\\hline
ABC  & Always Best Connected  \\\hline
AMPS  & Advanced Mobile Phone System \\\hline
AP   & Access Point  \\\hline
BS   & Base Station \\\hline
EDGE & Enhanced Data rates for GSM Evolution \\\hline
GPRS  & General Packet Radio Service \\\hline
GRA  & Grey Relational Analysis \\\hline
GSM   & Global System for Mobile communication \\\hline
HNE   & Handover Necessity Estimation \\\hline
HTCE   & Handover Triggering Condition Estimation \\\hline
HTS   & Handover Target Selection \\\hline
IEEE 802.11 & WiFi standard \\\hline
IEEE 802.16 & WiMAX standard \\\hline
IEEE 802.21 & MIH standard \\\hline
IMT-2000 & International Mobile Telecommunications-2000 \\\hline
IP & Internet Protocol \\
\end{tcolorbox}
\end{table}

\newcolumntype{Y}[1]{>{\raggedright\arraybackslash}p{#1}}
\begin{table}[H]
\begin{tcolorbox}[tab2,tabularx={X||p{9cm}}]

\begin{flushleft}
\textbf{\Large{\textit{Acronym Contd.}}}
\end{flushleft}

 &
\begin{flushleft}
\textbf{\Large{}}
\end{flushleft}
      \\\hline\hline
JTACS & Japanese Total Access Communication System \\\hline
LTE &  Long Term Evolution \\\hline
MICS & Media Independent Command Service \\\hline
MIES & Media Independent Event Service \\\hline
MIH  & Media Independent Handover\\\hline
MIIS & Media Independent Information Service \\\hline
MISAP & MIH Service Access Point \\\hline
MN   & Mobile Node \\\hline
QoS   & Quality of Service \\\hline
RSS   & Received Signal Strength \\\hline
TACS  & Total Access Communication System \\\hline
UMTS  & Universal Mobile Telecommunications Service \\\hline
VHO & Vertical Handover \\\hline
VoIP  & Voice over IP \\\hline
WCDMA  & Wideband Code Division Multiple Access \\\hline
WiFi  & Wireless Fidelity \\\hline
WiMAX  & Worldwide Interoperability for Microwave Access \\\hline
WLAN  & Wireless Local Area Network \\\hline
WMAN & Wireless Metropolitan Area Network \\
\end{tcolorbox}
\end{table}

%
%
%
%
%
%
%

\clearpage

%% file: chap1.tex

\chapter{Introduction}
\label{chp:1}
\newpage
\section{Introduction}
{\Huge $\mathbb{W}$}ith rapid growth in the use of the internet and wireless services, the challenge to support generalized mobility and provision of ubiquitous services to users while integrating diverse access technologies (GSM, 3G, 4G, WLAN, WiMAX and Bluetooth), has attracted research attention. Due to increased demand for mobile data, users now require access networks that use multiple layers (macro as well as micro cells), and multiple technologies to meet growing needs. As a mobile node (MN) moves within a heterogenous environment, satisfactory quality of service (QoS) is desired by ensuring efficient vertical handover.

Vertical handover can be defined as when an MN moves from one access network to another while maintaining the live call or session. In contrast to horizontal handovers, vertical handovers can be instigated for convenience rather than connectivity purposes \cite{RefJ6}. As such, the choice to perform vertical handover may depend on factors such as available bandwidth, received signal strength (RSS), access cost, dwell time, security, speed, etc. \cite{RefJ6,RefJ1,RefJ2,RefJ3,RefJ4,RefJ5}. For optimal decision making, it is imperative to weigh the benefits against the detriments before initiating a vertical handover.
\subsection{Wireless Mobility}
\begin{figure*}

 \begin{center}
 \includegraphics[scale=0.5]{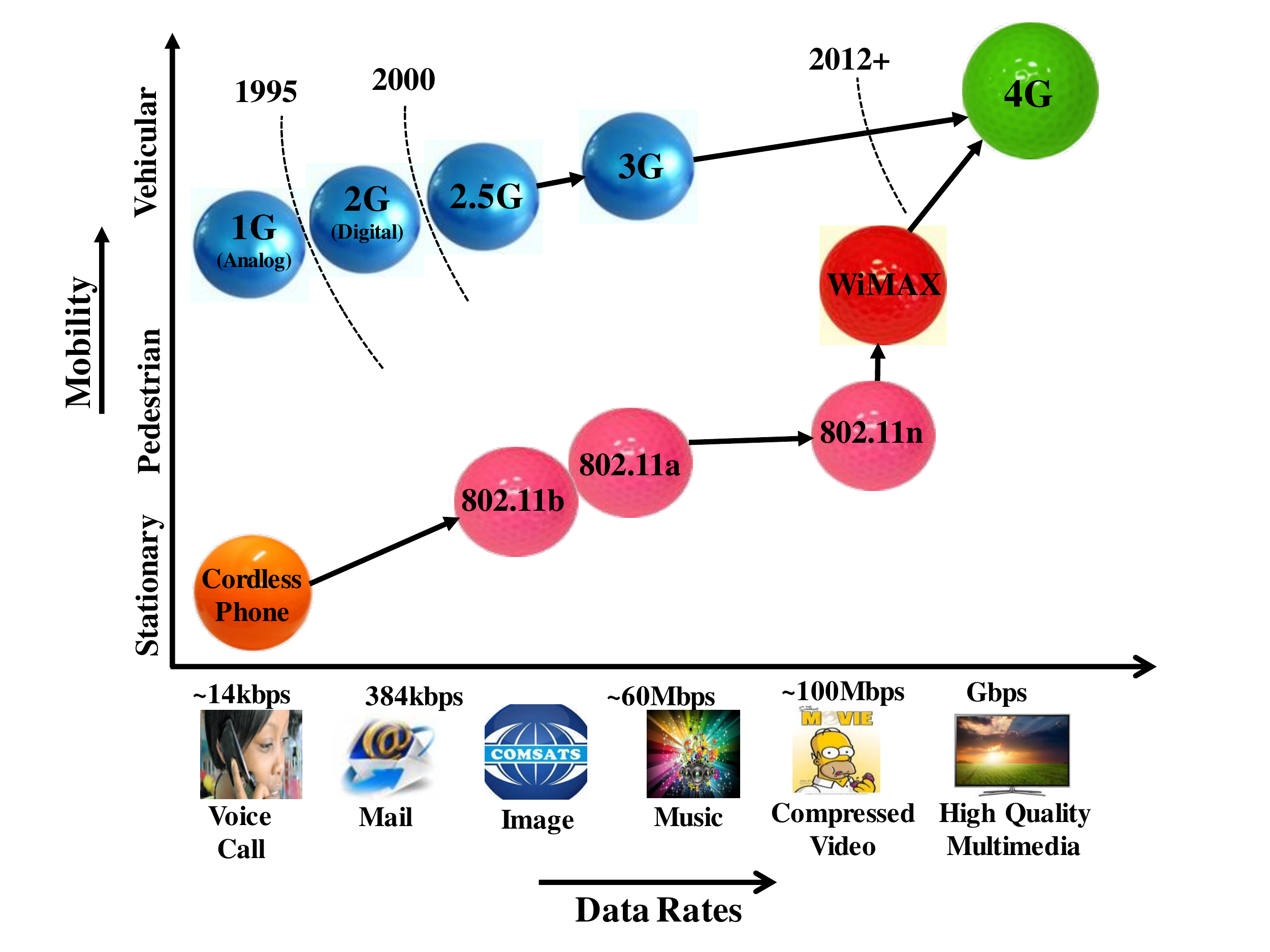} 
 \end{center}
\caption{Evolution of Wireless Communication}
\label{fig:1a}       
\end{figure*}

Wireless technologies have become a fundamental part of people’s day-to-day life. Driven by growing demands, wireless communication has evolved from the first to the fourth generation. As the internet and other bandwidth-hungry services become more prolific in the society, reliable services (voice, data and video) in a highly mobile environment are set goals \cite{RefB4}. Ultimately, wireless mobility systems being deployed in a network need to meet business objectives. This section presents a brief sampling of the immense array of wireless mobility concepts and the evolving technologies. The evolution of wireless communication is shown in Fig. \ref{fig:1a}.
\newpage

\subsubsection{Cellular Technology Evolution}
Over the past 25 years, evolution of the internet and advances in wireless technologies have made remarkable impact on lives around the world. There has been substantial deployment of a host of wireless platforms that are centered at reducing
cost for operators as well as delivering new and value-added services to subscribers. Brief description of developments in the area of wireless communications and technologies are hereby presented.
\begin{description}
  \item[\textit{1G:}]  1G mobility systems were analog and proved to be a great advancement in communication mobility. Different 1G standards were used in various countries, such as Advanced Mobile Phone System (AMPS), Total Access Communication System (TACS), Japanese Total Access Communication System (JTACS) and Nordic Mobile Telephone (NMT) \cite{RefB4,RefJ15}. 1G was a major innovation in the telecommunication history. However, it was susceptible to the problems of quality of transmissions, security and wasteful utilization of spectrum and capacity \cite{RefB5}.

  \item[\textit{2G:}]  2G networks introduced digital circuit-switched technology which uses the spectrum in a more efficient way. 2G networks are presently serving the vast majority of mobile users and will remain in the market for a long time. The major 2G cellular standards are GSM, IS-136 and CdmaOne \cite{RefB5}.

  \item[\textit{2.5G:}] After 2G and before the 3G, a stepping-stone technology called Two and One-Half Generation (2.5G) was introduced. 2.5G is the realm of enhanced data services. The key 2.5G standards include General Packet Radio Service (GPRS), Enhanced Data rates for GSM Evolution (EDGE), CDMA2000 1xRTT and IS-95B. GPRS is an enhanced mobile data service for users of GSM and IS-136 \cite{RefB5}.

  \item[\textit{3G:}]  3G networks are characterized by higher data transmission speed, better system capacity and improved spectrum efficiency among other features \cite{RefJ15}. There is a range of technologies for 3G, all based around CDMA, including UMTS (with both FDD and TDD variants), CDMA2000 and Time Division-Synchronous Code Division Multiple Access (TD-SCDMA) \cite{RefB5}.

  \item[\textit{4G:}]  4G networks are also known as fourth-generation wireless presents broadband mobile communications that supersedes the third generation (3G) of wireless communications. Currently, only few countries in the world have tapped into its use. The 4G framework was proposed based on the key concept of integration. 4G services operate in names such as Long Term Evolution (LTE) and Ultra-Mobile Broadband (UMB)\cite{RefB5}.
      \begin{itemize}
        \item LTE is designed to provide higher data rates with over 100 Mbps for downlink and over 50 Mbps for uplink for every 20 MHz of spectrum, lower latency and packet-efficient system compared to 3G \cite{RefT2}. LTE also uses Orthogonal Frequency Division Multiple Access (OFDMA) for the downlink and Single Carrier Frequency Division Multiple Access (SC-FDMA) for the uplink and employs Multiple-Input Multiple-Output (MIMO) with up to 4 antennas per station. LTE is designed to be all-IP based system and supports mobility and seamless service between heterogeneous wireless access networks \cite{RefT1}.
        \item Ultra Mobile Broadband is the successor to CDMA2000 EV-DO. UMB incorporates OFDMA, MIMO and Space Division Multiple Access (SDMA) cutting-edge antenna techniques to provide even better capacity, coverage and QoS. UMB can support peak download speeds as high as 280 Mbps in a mobile environment and over 75 Mbps for upstream transmission with 4x4 MIMO configuration \cite{RefT2}.
      \end{itemize}
\end{description}

\subsubsection{Mobile Broadband Wireless Technology Evolution}
Wireless broadband communication is the marriage of the two notably growing sectors in recent years: broadband communication and wireless mobile communication. During the same period, internet has been evolving from a curious academic tool to having about a billion users\cite{RefJ17}. Parallel to the growth of internet, the development of broadband technology has been accelerated to offer high-speed internet access.
\begin{description}
  \item[\textit{WiFi:}]  WiFi is a wireless LAN based on the IEEE 802.11 family of standards enhanced to support higher data rates and provide better QoS. It is primarily a WLAN technology designed to provide in-building broadband coverage. This standard operates in the unlicensed 2.4GHz and 5GHz band. The standards includes 802.11b, 802.11a, 802.11i, 802.11e, 802.11g, 802.11n, \cite{RefB4} etc. WiFi has become a defacto standard for broadband connection in homes, offices, public hot-spots and educational environments. In the past couple of years, a significant number of municipalities and local communities around the world have taken the initiative to get WiFi systems deployed in outdoor to provide broadband access to city centers as well as to rural and under-served areas.\\

  \item[\textit{WiMAX:}]  WiMAX is designed to accommodate both fixed and mobile broadband applications. It is based on the IEEE 802.16 standard and focuses on last-mile applications of wireless technology for broadband access \cite{RefB4}. However, WiMAX is different from WLAN and wireless mobility systems like GSM, CDMA and UMTS. It is unique in the sense that it provides broadband access to multiple users in the same geographical area. It uses microwave radios as its fundamental transport medium, making it adaptable to older technologies.
\end{description}
\subsection{Convergence of Heterogeneous Wireless Networks}
The mobile wireless community is upgrading and deploying a host of wireless platforms that are centered at reducing cost for operators as well as provide new and improved services to subscribers. There are several benefits for the subscriber and operator when two or more technologies are combined. However, a major concern is likely to be seamless mobility over heterogeneous wireless networks. 3G/WiFi inter-operability has been a focus of the wireless industry and this is evident through the use of Mobile IP. Mobile IP facilitates the continuity of data session when the user is mobile, by facilitating handovers between different access networks. The objective of convergence is to provide more services that will be purchased and used by customers \cite{RefB4}. Thus, the integration of wireless mobility with 802.11 enables the possibility of numerous service provision.

To some extent, both 3G and WiFi services are complimentary, with each having its unique benefits. WiFi can be used to provide higher data rates, while 3G data service provides greater roaming and voice service on a global basis. Convergence will provide mobile users with an Always Best Connected (ABC) feature, high security, available at any time, affordable cost, one billing, low latency and high QoS broadband experience \cite{RefB4}. The ABC means providing seamless service across several wireless access networks and an optimum service delivery through the best available network \cite{RefJ19}. Convergence is leading the way to the future.

\subsection{Horizontal vs. Vertical Handover}
\begin{figure*}
 \begin{center}
 \includegraphics[scale=0.3]{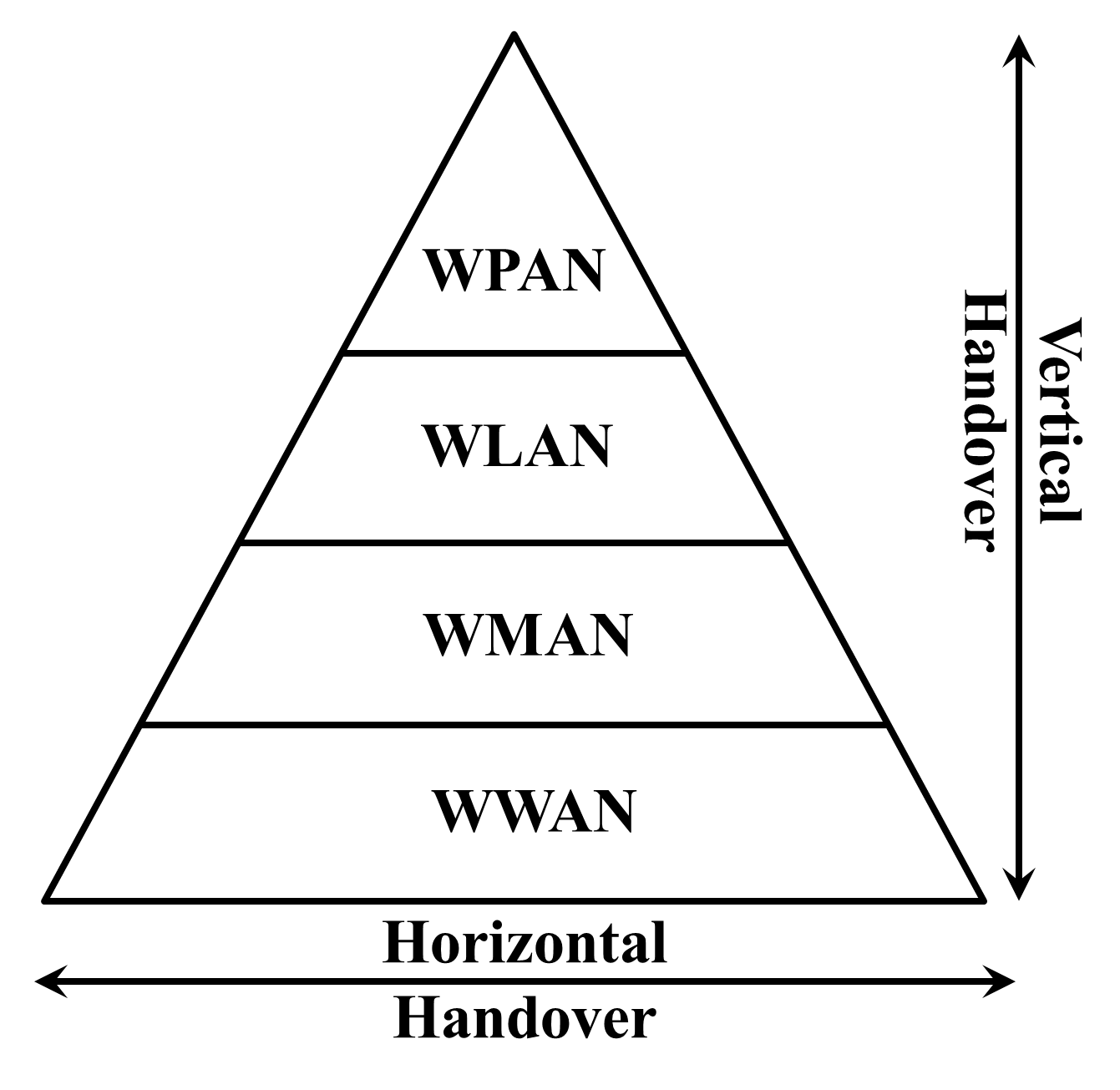} 
 \end{center}
\caption{Horizontal vs. Vertical Handover}
\label{fig:1b}       
\end{figure*}

Horizontal handover occurs when an MN moves between two coverage cells using the same technology. It is also known as Intra-cell (Intra-domain) handover \cite{RefT2} and running services are sustained by masquerading the change of IP address like in Mobile IP \cite{RefJ14}. Vertical handover on the other hand is a handover between two different access technologies. It is also known as Inter-cell (Inter-domain) handover \cite{RefT2}, which occurs when the MN moves into an adjacent cell or coverage area and all connections are transferred to a new BS or AP \cite{RefJ14}. Fig. \ref{fig:1b} shows the scope of operation between horizontal and vertical handovers. Considering the goals of Next Generation Networks (NGN), much attention will be given to vertical handovers in near future. The focus of this research will be on implementing a vertical handover decision making scheme with consideration for fading in a heterogeneous wireless environment.
\section{Scope and Methodology}

Several vertical handover algorithms have been proposed to initiate handover at the optimal time to the optimal network based on various network metrics \cite{RefJ6,RefJ1,RefJ2,RefJ3,RefJ4,RefJ5,RefJ7,RefJ8,RefJ9,RefJ10}. A comprehensive survey of related schemes can be found in chapter 2 of this thesis.
These vertical handover schemes have either failed to consider the effects of fading in handover design or the algorithms are impractical to be implemented.

The study will perform analysis and implementation of the proposed vertical handover scheme considering the following wireless networks: WiMAX, 3G and WLAN. The study will focus on considering the effects of shadow fading phenomenon in making optimal handover decisions. The goal will be achieved through the following methodology:
    \begin{itemize}
      \item Reproduction of the results of existing works
      \item Enhancing the geometrical models for HNE and HTCE by considering fading effects
      \item HTS, which is an important decision required prior to handover will be considered
      \item Examination and analysis of the enhanced model
      \item Comparison of earlier existing models with the proposed one
      \item Analysis of the results
    \end{itemize}

\section{Motivation}\label{ch1:m}

Due to alarming growth in the number of wireless users, the demand for acceptable QoS levels has increased. An efficient vertical handover plan must exist to accommodate the increase in demand for data traffic and also provide support for high mobility. It is unrealistic for any technology to provide support for high bandwidth, fast mobility, low latency and wide-area data service simultaneously to a large number of users at the same time \cite{RefJ11, RefJ14}.

Since there is no technology that could offer ubiquitous coverage, the motivation of this study is implement a vertical handover scheme that will address the issues of detrimental handovers and maximize user satisfaction. Every system has its merits and demerits, and no single technology out-performs other existing technologies till date \cite{RefJ11}. In order to maintain uninterrupted connectivity and get best services at all times in a heterogenous environment, it is required for an MN to switch connections among available access networks when the need arise.
The proposed scheme can be integrated into the existing MIH standard of the IEEE 802.21 protocol.

\section{Statement of Problem}
RSS-based algorithms are easy to implement, however, these algorithms are seriously limited by slow fading \cite{RefJ8}. Due to fading, it is almost impossible to accurately predict the RSS at specific locations \cite{RefB6}. Shadowing complicates cellular planning, and as such, previous works have neglected its effect in pre-handover decision making. The study presents an amoebic based geometric model that extends the ideal circular based models employed in previous works by considering the effect of fading.

Significant research efforts are being made to enable optimal vertical handover, but the task of enabling seamless mobility across diverse access networks is extremely challenging. To achieve this, not only handover execution, but pre-handover decisions (HNE, HTCE and HTS) are important. The existing works are limited to optimization of these pre-handover decisions with the assumption that the coverage region is perfectly circular. The study will extend the existing models by considering fading effects and aims to optimize HNE, HTCE and HTS in a realistic scenario.

\section{Thesis Organization}

The thesis is organized as follows: A detailed review of existing works related to vertical handovers is presented in the research literature of Chapter \ref{chp:2}. In Chapter \ref{chp:3}, an overall framework of the proposed handover scheme, as well as details of the three key components, Handover Necessity Estimation (HNE), Handover Target Selection (HTS) and Handover Triggering Condition Estimation (HTCE) are implemented. Theoretical and simulation based experimental results of the proposed scheme are presented and discussed in Chapter \ref{chp:4}. Conclusions of the research work, and suggestions for future research directions are given in Chapter \ref{chp:5}.
\clearpage

%% file: chap2.tex

\chapter{Review of Related Literature}
\label{chp:2}
\newpage
\section{Introduction}
{\Huge $\mathbb{R}$}elated works on vertical handover can be grouped into Three types \cite{RefJ5}, the RSS-based, Cost-based and Other related works which are briefly described.\begin{description}
                                    \item[\textit{\textbf{RSS-based related works}}:] Several RSS-based handover algorithms have been developed for wireless communications. A novel algorithm was developed using the concept of dynamic boundary area to support seamless vertical handover between the 3G and WLAN in \cite{RefJ4}. A traveling distance prediction based handover decision method \cite{RefJ2}, dwell time prediction model \cite{RefJ3} and a linear approximation of the travelled distance \cite{RefJ1} were proposed to minimize the probability of unnecessary handover. However, the geometric models considered were not of a realistic coverage cell shape.
                                    \item[\textit{\textbf{Cost-based related works}}:] Handover cost is a function of the available bandwidth, security, power consumption and the monetery cost \cite{RefJ11}. As the need for voice and video services rise, available bandwidth, power consumption, security, etc., will be a major factor used to indicate network conditions to trigger handovers. In \cite{RefJ9}, available bandwidth and monetary cost were used as metrics for handover decisions. Cost-based algorithms are usually complex as they require collecting and normalization of different network metrics.
                                    \item[\textit{\textbf{Other related works}}:] An analytical framework to evaluate vertical handover algorithms with new extensions for traditional hysteresis based and dwell timer-based algorithms was proposed in \cite{RefJ10}. Using probability approach, \cite{RefJ7} worked on the assessment of a Wrong decision probability (WDP), which assures a trade-off between network performance maximization and mitigation of the ping-pong effect. The proposed algorithm was able to reduce the vertical handover frequency and keep the received bits as high as possible.
                                  \end{description}

The focus of this work is to introduce an amoebic based geometric model that extends the ideal circular coverage model employed in previous works. The work considers the RSS-based dwell time approach. RSS-based algorithms are easy to implement, however, these algorithms are seriously limited by slow fading \cite{RefJ8}. Slow fading can be caused by events such as shadowing, where a large obstruction such as a hill or large building obscures the main signal path between the transmitter and the receiver. This work presents a novel and realistic model that depicts the actual behaviour of a WLAN coverage area, considering the effects of fading. The proposed model will ensure an efficient handover decision considering the following factors:\begin{itemize}
          \item The WLAN cell shape is not exactly circular, but irregular.
          \item The cell shape changes with changes in nature of obstruction at different instances, humidity, temperature etc \cite{RefB5}.
        \end{itemize}
\section{Amoebic Wireless Coverage Concept}\label{ch2:a}
Both theoretical and empirical propagation models show that average received signal power decreases logarithmically with distance.\footnote{~$PL(d)_{dB}= P(transmit)_{dB} - P(receive)_{dB} = PL(d_0) + 10\beta\log(d/d_0) + R_{\sigma}$, \\Where ~$R_{\sigma}$ is the Gaussian random variable with standard deviation, ~$\sigma$ and ~$\beta$ is the path loss exponent.} There exist a number of factors, apart from the frequency and the distance that influence losses encountered by propagated signals from the AP to the MN. In order to accurately model a wireless coverage area the factors that must be considered are:\begin{itemize}
                         \item the height of the MN antenna;
                         \item the height of the AP relative to the surrounding terrain;
                         \item the terrain irregularity (undulation or roughness);
                         \item the land usage in the surroundings of MN: urban, suburban, rural, open, etc.
                       \end{itemize}

Due to these effects, the coverage region does not remain circular, but of an irregular shape and this shape also changes with time. Thus, it is called an amoebic shaped coverage region. This paper presents an Amoebic WLAN cell which gives a realistic representation of the wireless coverage with a perspective of the shadowing concept. There are three different rates of variation as wireless signals are propagated away from the AP: we have (i) the very slow variation, called path loss, which is a function of distance between the AP and MN, (ii) slow variation, which results from shadowing effects, and (iii) fast variation, due to multi-path. Signal variations caused by multi-path, in the case where the direct signal is assumed to be totally blocked, are usually represented by a Rayleigh distribution\cite{RefB3}. The slow received signal variability due to shadowing is usually assumed to follow a Gaussian distribution~\cite{RefB3}.
\begin{figure*}
 \begin{center}
 \includegraphics[scale=0.6]{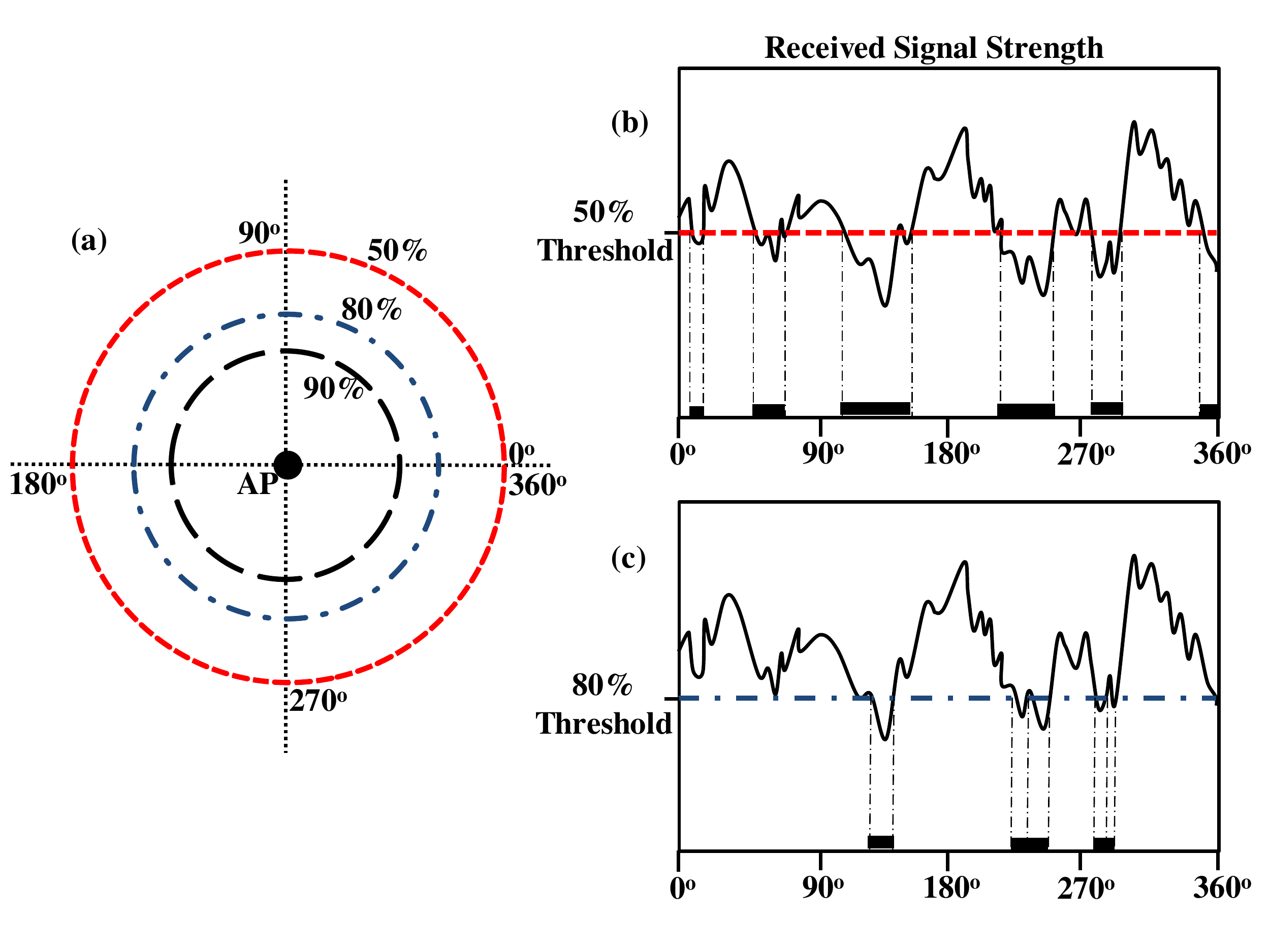} 
 \end{center}
\caption{(a) Coverage regions for 50\%, 80\% and 90\%. (b) Received Signal Strength (RSS) while moving along the 50\% contour. (c) RSS while moving along the 80\% contour. }
\label{fig:2a}       
\end{figure*}

This work considers the effect of shadow fading. Fig. \ref{fig:2a} gives a clearer picture of the behavior of wireless signals in a coverage area with coverage probabilities of 90\%, 80\% and 50\%. When an MN moves along contours as shown in Fig. \ref{fig:2a} (b) and (c), it may observe poor signal strength at some instance (or places) which can lead to bad speech quality (for voice telephony), high Bit Error Rate (BER) and low data rate (for data transmission) \cite{RefB6}. If the quality becomes too low for a longer duration of time, it may lead to termination of the connection.
These slow signal deviations due to shadowing follows a Gaussian distribution and is given as
\begin{equation}
\label{eqn:2a}
f(r) = \frac{1}{\sigma\sqrt{2\pi}}\exp\Big[{\frac{-(r-\mu)^{2}}{2\sigma^{2}}}\Big]
\end{equation}
Where ~$\mu$ is the mean value and ~$\sigma^{2}$ is the variance of the Gaussian random variable ~$r$.

Suppose we have the radius of the WLAN cell which is a continuous random variable R with PDF, ~$f(r)$, as shown in Equation (\ref{eqn:2a}) and we desire to evaluate the expectation ~$E[g(R)]$ for some function ~$g(r)$. This entails evaluating the integral,
\begin{equation}
\label{eqn:2b}
E[g(R)]=\int_{-\infty}^{\infty} f(r)g(r)\,dr
\end{equation}

Since the integral is not easily tractable by analytical or standard numerical methods, the study approached it by simulating realizations of ~$r_{1}$, ~$r_{2}$, ~$r_{3}$, … ~$r_{n}$ of ~$R$, and since the variance is finite, we apply the law of large numbers to obtain an approximation\cite{RefB3}.

\begin{equation}
\label{eqn:2c}
E[g(R)]  \sim \frac{1}{n} \sum_{i=1}^{n} g_{i}(r)
\end{equation}
The expression in Equation (\ref{eqn:2c}) gives justification for the Monte-Carlo simulation carried out in the study.
\section{Media Independent Handover (MIH) Architecture}

\begin{figure*}[h]

 \begin{center}
 \includegraphics[scale=0.6]{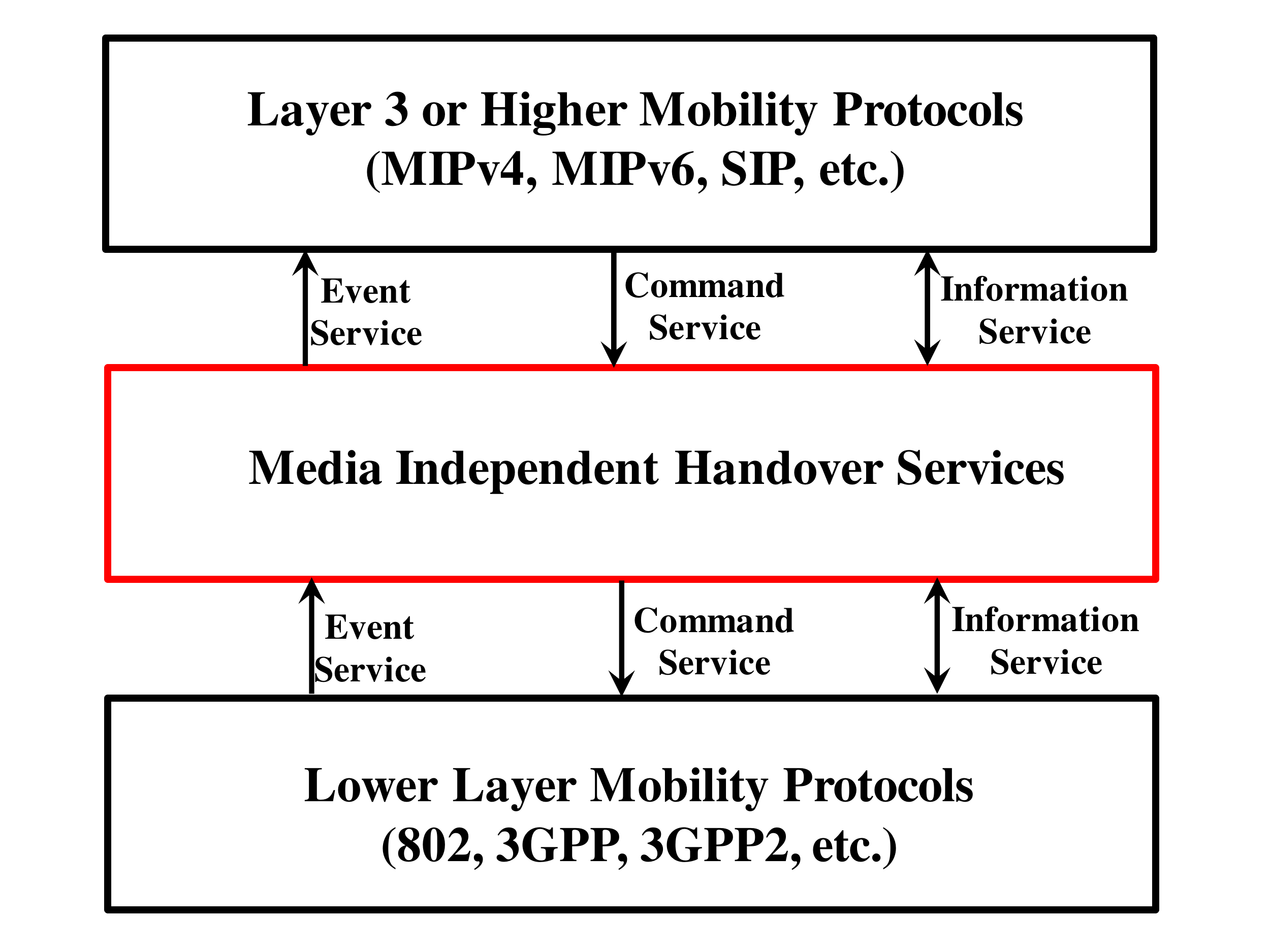} 
 \end{center}
\caption{Media Independent Handover Services}
\label{fig:2b}       
\end{figure*}

IEEE group proposed IEEE 802.21 standard, Media Independent Handover (MIH), to provide seamless vertical handover and desirable QoS requirements across heterogeneous network environments \cite{RefJ17}. This standard is intended to provide a generic interface between the link-layer users in the mobility-management protocol stack and existing media-specific link layers, such as those specified by 3rd Generation Partnership Project (3GPP), 3GPP2, and the IEEE 802 family of standards \cite{RefJ18}. However, in \cite{RefJ16}, limitations in MIH architecture were stated as follows:
\begin{itemize}
  \item The handover process is typically based on measurements and triggers initiated from link layers, which disregards the influence of the application and user preference information on mobility management.
  \item The network information provided by MIH lacks flexibility in the sense that only static and less dynamic information is derived.
\end{itemize}
This study aims to address the above problems by implementing a vertical handover scheme for optimal handover decisions which can be grafted into the MIH architecture. MIH provides three main services: Media Independent Event Service (MIES), Media Independent Command Service (MICS) and Media Independent Information Service (MIIS) \cite{RefJ18}, this is shown in Fig. \ref{fig:2b}
\subsection{Media Independent Event Service (MIES)}
The MIH Event Service (MIES) is responsible for communicating network critical events to upper layer mobility protocols. These events are used by the upper layers to determine optimal handover instant \cite{RefJ22}. It basically report events after detecting, e.g. connection establishment, broken links, imminent link breakdown \cite{RefJ17,RefJ18} etc. However, the MIH architecture is limited in providing specifications on the manner in which these events can be triggered. Usually
events are triggered when received signal strength (RSS) level falls below a predefined threshold, but the design of this threshold value becomes complex due to the effect of fading.

\subsection{Media Independent Information Service (MIIS)}
The Media Independent
Information Service (MIIS) provides a framework by which an MIH function, residing in the MN or in the network, discovers and acquires network information within a geographical area to expedite network selection and handovers \cite{RefJ22}. It is responsible for collecting all information required to identify if a handover is necessary or not and consequently provide them to the MN, e.g. available networks, locations, capabilities, cost \cite{RefJ17,RefJ18} etc.

\subsection{Media Independent Command Service (MICS)}
The MIH Command Service (MICS) enables higher layers to control the physical, data-link and logical-link layers \cite{RefJ22}. It is responsible for issuing the commands based on the information which is gathered by MIIS and MIES, e.g. MIH handover initiate, MIH handover prepare, MIH handover commit and MIH handover complete \cite{RefJ17,RefJ18}.

\section{Handover Management}
\begin{figure*}[h]

 \begin{center}
 \includegraphics[scale=0.6]{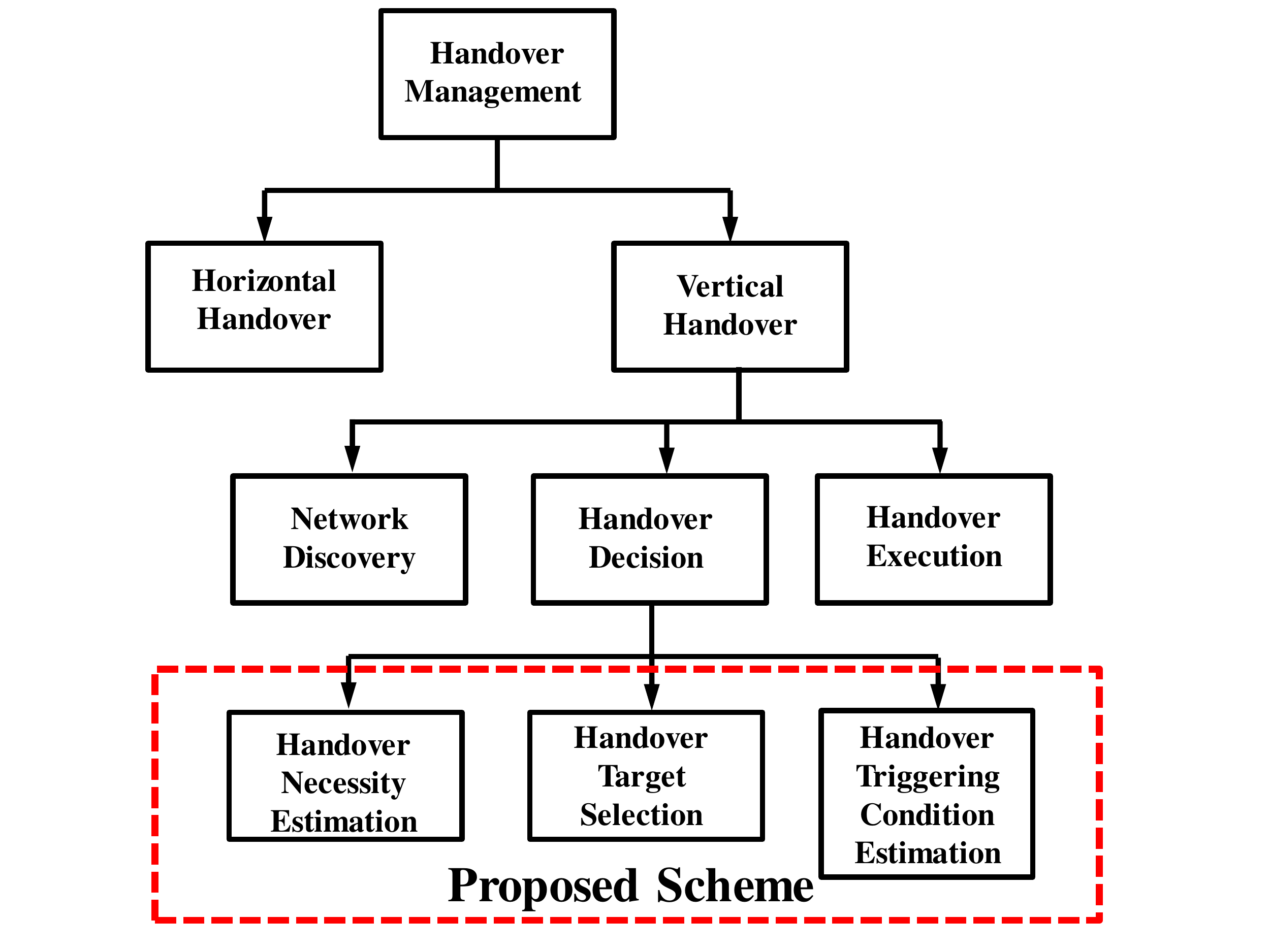} 
 \end{center}
\caption{Hierarchy of handover management in a heterogeneous network environment}
\label{fig:2c}       
\end{figure*}

There may be considerable overlap in adjacent coverage areas of wireless networks. Some overlap is desired in order to support handover when an MN is moving from one coverage area to another during a live call or session. Mobility management has enabled MNs to maintain their ongoing sessions particularly when moving between different access networks \cite{RefJ17}. Vertical handover is described as the handover between two access nodes of two different technologies, and since every technology has its unique mobility issue, there is need for efficient handover management in heterogeneous networks.  The hierarchy of handover management in a heterogeneous network environment is shown in Fig. \ref{fig:2c}. In this thesis, a vertical handover scheme is developed to provide seamless mobility and desirable QoS for mobile users.
\subsection{Handover Necessity Estimation (HNE)}
The study proposes a method to estimate the necessity for a handover. An amoebic based model that extends the ideal circle model employed in previous works \cite{RefJ1,RefJ2,RefJ3,RefJ4} for HNE is hereby proposed. Yan \textit{et al}. \cite{RefJ2} and Hussain \textit{et al}. \cite{RefJ3} employed the dwell time approach to develop models that kept the probabilty of handover failure and unnecessary handover within tolerable bounds, however, the model in \cite{RefJ2,RefT1} considered a bound of~$[0, 2\pi]$ which was wrong and the model in \cite{RefJ3} considered the angle of arrival and departure to lie within~$[0, \pi]$ bound, but it was impractical because it requires precise information (on the tangential angle of arrival of the MN) from the system.

The proposed algorithm calculates a time threshold based on various network parameters which include the handover failure or unnecessary handover probability information. The expression of handover failure or unnecessary handover probability is made by developing a mathematical model which assumes that the cell radius is stochastic and normally (Gaussian) distributed with defined mean and variance. The justification for proposing a normal distribution can be given in terms of the Central limit theorem \cite{RefB3}, as the total attenuation experienced in a wireless link results from the tallying of several individual shadowing processes forming a Gaussian distribution. The speed~$(v)$ of the MN and the direction of motion~$(\theta)$ are assumed to be uniformly distributed within bounds of~$[v_{min}, v_{max}]$ and~$[0, \pi]$, respectively. The predicted traveling time is compared against the time threshold and a handover is necessary only if the traveling time exceeds the predefined threshold value.
\subsection{Handover Triggering Condition Estimation (HTCE)}
This method is proposed to help in triggering a handover when the RSS of the current serving network is deteriorating. A lot of research attention has been on optimizing the handover triggering process to minimise connection breakdown as a mobile node (MN) traverses a heterogeneous wireless environment. The IEEE 802.21 working group proposed a MIH framework with an event service feature responsible for reporting the events after detecting, e.g. connection establishment,
broken links, imminent link breakdown \cite{RefJ18}etc. MIH defines the communication interface, however, it does not provide specifics on how events should be triggered \cite{RefJ18,RefJ22}. Mohanty S. \cite{RefJ4} first proposed the concept of dynamic boundary area to support seamless roaming between a 3G and WLAN coverage area. The model in \cite{RefJ4} was able to minimize handover failure by initiating a handover from a particular distance from the boundary of the WLAN coverage area, however, there was no consideration for fading in the wireless channel. Nguyen-Vuong Q. T. \cite{RefT2} presented an expression for the adaptive handover threshold, but suppressed the shadowing part of the model by passing it through a low-pass filter.

Abrar S. \textit{et al.}\cite{RefJ21} presented a model that was able to keep the probability of handover failure within tolerable bounds. The work \cite{RefJ21} slightly out-performed the work of Mohanty S. \cite{RefJ4}, however, the expression for the threshold triggering distance was not stated and the effect of fading was neglected in the model design. Yan X. \textit{et al.}\cite{RefT1} proposed a critical point for triggering a handover. The work in \cite{RefT1} presented a mechanism for managing a trade-off between connection breakdowns and WLAN usage while dynamically adapting to the speed of the MN. Erroneously, this work \cite{RefT1} considered a~$[0,2\pi]$ bound\footnote{We considered a~$[0,\pi]$ bound since there is only a possibility for the MN to move in any half part of the coverage cell} and also neglected the effect of fading. However, the geometric models considered were not of a realistic coverage cell shape. The proposed HTCE will be able to keep the connection breakdown probability below desirable limits, and provides the user with control over the trade-off between connection breakdown probability and WLAN usage.

\subsection{Handover Target Selection (HTS)}
Selecting an optimal network to perform handover in a heterogeneous environment is a complex task. There are lots of benefits (both to users and operators) which is derived from the optimization process, some of which include: improved QoS, minimizing unnecessary handovers, load balancing, congestion avoidance etc. Network selection is the process by which an MN or a network entity selects an available network to establish network-layer connectivity \cite{RefJ18}. Savitha K. \textit{et al.} \cite{RefJ20} used Simple additive weighting method (SAW) and Weighted product model (WPM) to choose the best network from the available access networks, however, only bandwidth, delay, jitter and cost were used as performance metrics. Omoniwa B. \cite{RefJ13} used Grey relational analysis (GRA) to effectively solve a multi criteria decision problem, however, the focus was on selecting an optimal robot based on several performance attributes.

In this work, GRA is employed because of its computational ease and convergence speed in arriving at an optimal decision. In addition, a new performance metric (ie. the estimated dwell time of the MN) is used alongside other performance attributes to select the optimal access network.
\subsubsection{Multi Criteria Target Network Selection (MCTNS) using GRA}
Grey system theory was introduced by Deng Ju-Long in 1982 \cite{RefJ13}. Grey system theory is built on the notion that a system is uncertain, and that the information contained in the system is inadequate to construct a reliable model that describes the system \cite{RefJ12}. This suggests that grey models are appropriate for predicting future events where system designers can make use of very limited data. This study will lay emphasis on the GRA for selecting an optimal target network.

The GRA can be clearly broken down into four steps, namely, grey relational generation, a reference sequence generation, grey relational coefficient calculation and grey relational grade calculation. These steps are further explained:
\begin{description}
  \item[\textit{Grey relational generating (GRG):}] GRG is a normalization process where all performance attributes are processed into a comparable sequence. Equation (\ref{eqn:2d1}) is used to normalize the higher the better attributes, Equation (\ref{eqn:2d2}) for lower the better and for the closer to the desired the better attributes, Equation (\ref{eqn:2d3}) is used for normalization. For MCTNS problems presented in this work, m is given as the network alternatives and n as performance attributes. Given a target network selection problem, ~$Y_{i} = \{y_{i1}, y_{i2}, ..., y_{ij}, ..., y_{in}\},$ we can deduce the comparability sequence, ~$X_{i} = \{x_{i1}, x_{i2},..., x_{ij},..., x_{in}\},$ for all ~$i= 1, 2, ..., m$ and ~$j= 1, 2, ..., n$.
      \begin{equation}
      \label{eqn:2d1}
      x_{ij} =  \frac{y_{ij} - Min\{y_{ij}, i= 1, 2,..., m\}}{Max\{y_{ij}, i= 1, 2,..., m\}- Min\{y_{ij}, i= 1, 2,..., m\}}
      \end{equation}
      \begin{equation}
      \label{eqn:2d2}
      x_{ij} =  \frac{Max\{y_{ij}, i= 1, 2,..., m\} - y_{ij}}{Max\{y_{ij}, i= 1, 2,..., m\}- Min\{y_{ij}, i= 1, 2,..., m\}}
      \end{equation}
      \begin{equation}
      \label{eqn:2d3}
      x_{ij} = 1 - \frac{|y_{ij} - y_{j}|}{Max\Big\{Max\{y_{ij}, i= 1, 2,..., m\} - y_{j}, y_{j} - Min\{y_{ij}, i= 1, 2,..., m\}\Big\}}
      \end{equation}

  \item[\textit{Reference sequence generation (RSG):}] After the normalization process using GRG, all performance values are defined within the range~$[0, 1]$. If the value~$x_{ij}$ with an attribute ~$j$ of access network alternative,~$i$, which equals 1 or approaches 1, it implies that the performance of alternative ~$i$ is the most suitable for attribute,~$j$ \cite{RefJ13}. The reference sequence is given as~$X_0 = \{x_{01}, x_{02}, …, x_{0j}, …, x_{0n}\}$ and the study sets the sequence as all ones~$\{1, 1, …, 1, …, 1\}$ with the aim of finding the alternative whose~$X_i$ is closest to~$X_0$.

  \item[\textit{Grey relational coefficient calculation (GRC):}] To determine how close the comparability sequence is to the reference sequence, we calculate the GRC as shown in Equation (\ref{eqn:2d4}). The role of the distinguishing coefficient,~$\zeta$ expands and compresses the range of GRC.
      \begin{equation}
      \label{eqn:2d4}
      \gamma(x_{0j}, x_{ij}) = \frac{\Delta_{min} + \zeta \Delta_{max}}{\Delta_{ij} + \zeta\Delta_{max}}
      \end{equation}
      Where ~$\gamma(x_{0j}, x_{ij})$ is the GRC between~$x_{0j}$ and ~$x_{ij}$,\\
      ~$\Delta_{ij} = |x_{0j} - x_{ij}|$,\\
      ~$\Delta_{min} = Min\{\Delta_{ij}, i= 1, 2,..., m; j= 1, 2,..., n\}$,\\
      ~$\Delta_{max} = Max\{\Delta_{ij}, i= 1, 2,..., m; j= 1, 2,..., n\}$

  \item[\textit{Grey relational grade calculation (GRGC):}] After the grey relational coefficient is calculated, the grey relational grade between the~$X_i$ and~$X_0$ is given as,
      \begin{equation}
      \label{eqn:2d5}
      \Gamma(X_{0}, X_{i}) = \sum_{j=1}^{n} w_{j}\gamma(x_{0j}, x_{ij})
      \end{equation}

      Where~$\Gamma(X_{0}, X_{i})$ is the GRG between~$x_{0j}$ and~$x_{ij}$, the weight of attribute,~$j$ is expressed as, ~$w_j$ and it is subject to the system designers' view of a particular problem. An alternative that has the closest value to the reference value is ranked best.

\end{description}
The study implements the GRA approach by presenting two case studies in Chapter \ref{chp:3}. This approach is chosen because it is mathematically comprehensible and computationally faster than other multi criteria selection algorithms \cite{RefJ13}.
\section{Summary}

In this chapter, a comprehensive survey of existing vertical handover schemes was presented. A brief overview of the MIH Architecture and the GRA algorithm were also presented. These works were classified into three categories: RSS-based, cost-based and other work based. Existing research literatures failed to consider the effect of fading. This research work focuses on this issue and provides an integrated solution to enable optimal vertical handover decision. In the next chapter, the framework of the proposed scheme is presented.

%% file: chap3.tex
\chapter{Proposed Scheme}
\label{chp:3}
\newpage
\section{Introduction}
{\Huge $\mathbb{T}$}his chapter presents a novel vertical handover scheme, comprised of three components: Handover Necessity Estimation (HNE), Handover Target Selection (HTS) and Handover Triggering Condition Estimation (HTCE). All parameters in the models for HNE and HTCE were derived from extensive geometric and probability analysis, while Grey relational analysis (GRA) was used for HTS. The proposed approach correctly simulates the actual behavior of the MN traversing a heterogeneous wireless environment.
\section{Handover Necessity Estimation}
\begin{figure*}[h]
\begin{center}
 \includegraphics[width=0.9\textwidth]{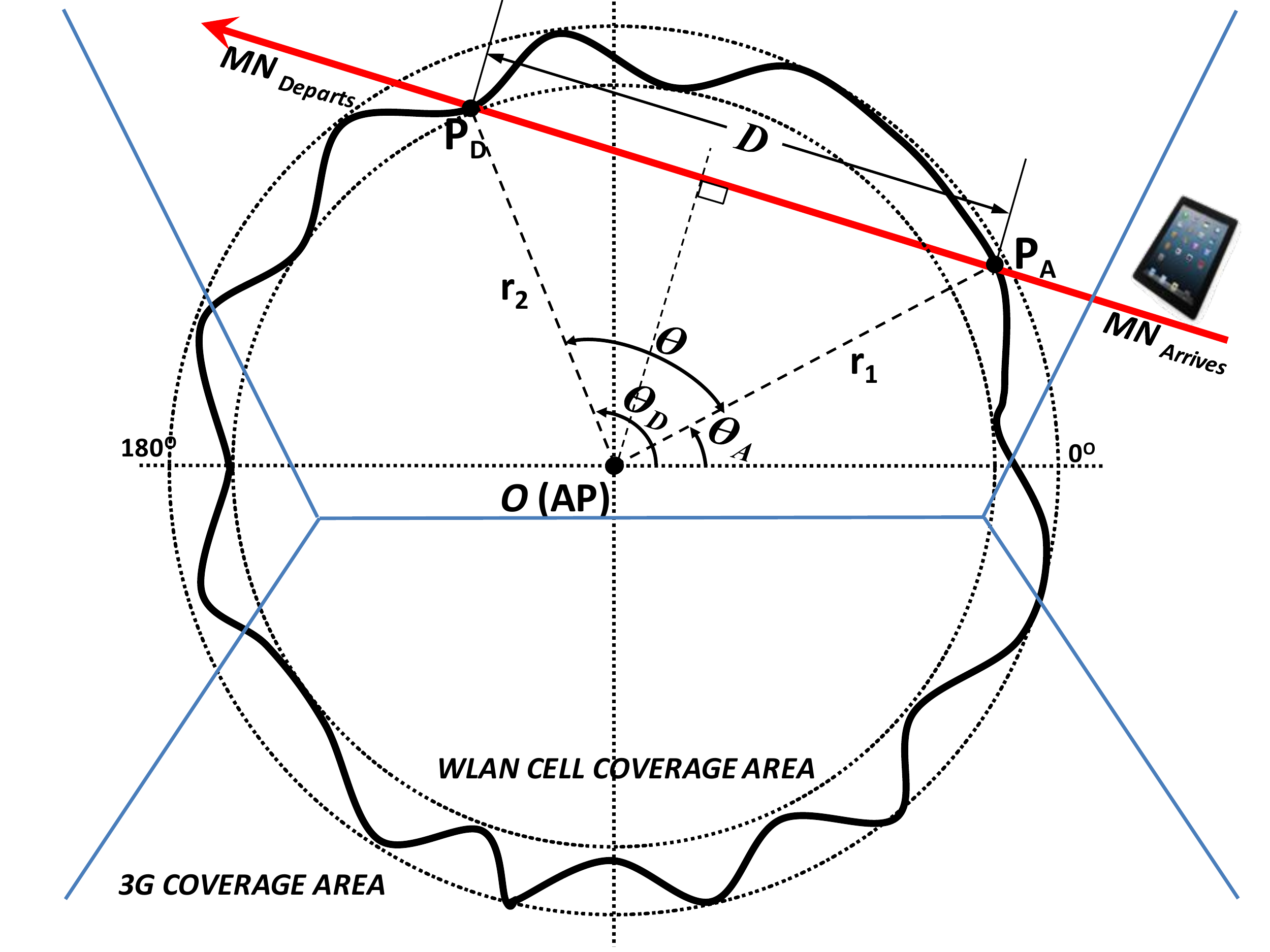}
\caption{A mobile node entering an amoebic WLAN coverage area.}
\label{fig:3a}       
\end{center}
\end{figure*}

The work assumes that when an MN is in the coverage area of it's present access network, adjacent to a WLAN cell,\footnote{We present an amoebic based model that extends the ideal circle model employed in previous works \cite{RefJ1,RefJ2,RefJ3,RefJ4}} it may enter the boundary area of the WLAN cell at any point, ~$P_A$ and move along the path ~$|P_AP_D|$, making exit from any point, ~$P_D$ on the coverage boundary (as shown in Fig. \ref{fig:3a}). We further assume that the speed,~$v$ of the MN is uniformly distributed in ~$[v_{min}, v_{max}]$. The cell radius is assumed to be stochastic and normally (Gaussian) distributed with defined mean and variance. The justification for having a normal distribution can be given in terms of the Central limit theorem\footnote{The central limit theorem states that given a distribution with mean, ~$\mu$ and variance, ~$\sigma^2$, the sampling distribution of the mean approaches a Gaussian distribution with mean, ~$\mu$ and variance, ~$\frac{\sigma^2}{n}$, where ~$n$ is the number of samples.} \cite{RefB3}, as the total attenuation experienced in a wireless link results from the tallying of several individual shadowing processes forming a Gaussian distribution.

The angle of arrival, ~$\Theta_A$ and angle of departure, ~$\Theta_D$ are assumed to be uniformly distributed within the bound of ~$\Theta$\footnote{We considered ~$[0,\pi]$, while ~$[0,2\pi]$ \cite{RefJ2} and ~$[0,\pi]$ \cite{RefJ3} with angle of arrival with respect to tangential line within ~$[0,\frac{\pi}{2}]$ were considered in previous works.} and we express the angle between random positions of ~$P_A$ and ~$P_D$ as ~$\Theta = | \Theta_D - \Theta_A|$. A realistic coverage area of the WLAN cell with an amoebic structure is considered in this work. As there is only a possibility that the MN moves in and out of the coverage area in any half section of Fig. \ref{fig:3a}, we therefore derive an expression to calculate the probability distribution function (PDF) of ~$\Theta$. The PDF of the arrival and departure of the MN from the WLAN coverage at point ~$P_A$ and ~$P_D$ respectively is given by

\begin{equation}
\label{eqn:4}
f_{\Theta_A}(\theta_{A})=\left\{
\begin{array}{c l}
    \frac{1}{\pi}, & 0\leq\Theta_{A}\leq\pi,\\
    0, & Otherwise.
\end{array}\right.
\end{equation}

\begin{equation}
\label{eqn:5}
f_{\Theta_D}(\theta_{D})=\left\{
\begin{array}{c l}
    \frac{1}{\pi}, & 0\leq\Theta_{D}\leq\pi,\\
    0, & Otherwise.
\end{array}\right.
\end{equation}
Since the arrival and departure points of the mobile nodes are independent, the Joint PDF is therefore given as product of their individual marginal functions.

\begin{equation}
\label{eqn:6}
f_{\Theta_A, \Theta_D}(\theta_{A}, \theta_{D})=\left\{
\begin{array}{c l}
    \frac{1}{\pi^{2}}, & 0\leq\Theta_{A}, \Theta_{D}\leq\pi,\\
    0, & Otherwise.
\end{array}\right.
\end{equation}
We find the cumulative distribution function (CDF) of ~$\Theta$ by

\begin{equation}
\label{eqn:7}
\begin{split}
 F_{\Theta}(\theta)&=P(\Theta\leq\theta)\\
    &= \int\int_{\epsilon}f_{\Theta_A, \Theta_D}(\theta_{A}, \theta_{D})\,d\Theta_{D} d\Theta_{A}
\end{split}
\end{equation}

Where ~$\epsilon$ is a set of arrival and departure points along the coverage boundary for the MN such that ~$ 0 \leq \Theta \leq \pi$. ~$P (\Theta\leq\theta) = 0$ for ~$\theta < 0$ and ~$P (\Theta\leq\theta) = 1$ for ~$\theta > \pi$ \cite{RefJ2}. From Fig. \ref{fig:3a}, Equation (\ref{eqn:7}) can be expressed as

\begin{equation}
\label{eqn:8}
\begin{split}
 F_{\Theta}(\theta)&=\frac{1}{\pi^{2}}\Bigg(\int_{0}^{\theta}\int_{0}^{\theta + \theta_{D}}d\Theta_{D} d\Theta_{A} + \int_{\theta}^{\pi-\theta}\int_{\theta_{D} - \theta}^{\theta_{D} + \theta}d\Theta_{D} d\Theta_{A}  \\
    &+\int_{\pi - \theta}^{\pi}\int_{\theta_{D} - \theta}^{\pi}d\Theta_{D} d\Theta_{A}\Bigg)
\end{split}
\end{equation}

The final expression of CDF is obtained as:

\begin{equation}
\label{eqn:9}
F_{\Theta}(\theta)= \frac{(2\pi - \theta)\theta}{\pi^{2}}, 0\leq\Theta\leq\pi
\end{equation}

The corresponding PDF of ~$\Theta$ is given by:

\begin{equation}
\label{eqn:10}
f_{\Theta}(\theta)=\left\{
\begin{array}{c l}
    \frac{2(\pi-\theta)}{\pi^{2}}, &   0\leq\Theta\leq\pi,\\
    0, &   Otherwise.
\end{array}\right.
\end{equation}
We can now use the PDF of ~$\Theta$ to compute the PDF of the traversing time by the MN, ~$t_{WLAN}$. Using the Cosine formula, we formulate a geometric expression of the traversing distance, ~$D$ from Fig. \ref{fig:3a}
\begin{equation}
\label{eqn:11}
D = \sqrt{r_{1}^{2} +r_{2}^{2} - 2r_{1}r_{2}\cos\theta}
\end{equation}
The traversal distance through the WLAN cell, ~$D$ depends on the traversing angle ~$\theta$.
\begin{equation}
\label{eqn:12}
\begin{split}
 t_{WLAN}&=g(\theta)\\
    &= \frac{\sqrt{r_{1}^{2} +r_{2}^{2} - 2r_{1}r_{2}\cos\theta}}{v}
\end{split}
\end{equation}
Where, $r_{1}$ and $r_{2}$ are the distances of the MN from the access point at the time of entry and exit from the coverage region respectively.

The PDF of the traversing time can thus be expressed as \cite{RefB1}
\begin{equation}
\label{eqn:13}
F(T) = \sum_{i=1}^{n} \biggl|\frac{f(\theta_{i})}{g'(\theta_{i})}\biggl| _{\theta_{i} = g^{-1}(T)}
\end{equation}
Where ~$\theta$ is the root of function ~$g(\theta)$, and ~$g'(\theta)$ is the derivative of ~$g(\theta)$.
\begin{equation}
\label{eqn:14}
\theta = \arccos\Big(\frac{r_{1}^{2} +r_{2}^{2} - t_{WLAN}^{2}v^{2} }{2r_{1}r_{2}}\Big)
\end{equation}
We have the derivative of ~$g(\theta)$ as
\begin{equation}
\label{eqn:15}
g'(\theta) = \frac{r_{1}r_{2}\sin\theta}{v\sqrt{r_{1}^{2} +r_{2}^{2} - 2r_{1}r_{2}\cos\theta}}
\end{equation}
Thus, substituting the Equation (\ref{eqn:14}) into (\ref{eqn:15}) to get,
\begin{equation}
\label{eqn:16}
\begin{split}
 g'(\theta)&=\frac{r_{1}r_{2}\sin\Big(\arccos(\frac{r_{1}^{2} +r_{2}^{2} - t_{WLAN}^{2}v^{2} }{2r_{1}r_{2}})\Big)}{v\sqrt{r_{1}^{2} +r_{2}^{2} - 2r_{1}r_{2}\cos\Big(\arccos(\frac{r_{1}^{2} +r_{2}^{2} - t_{WLAN}^{2}v^{2} }{2r_{1}r_{2}})\Big)}}\\
    &= \frac{\sqrt{4r_{1}^{2}r_{2}^{2} - (r_{1}^{2} +r_{2}^{2} - t_{WLAN}^{2}v^{2})^{2}}}{2t_{WLAN}v^{2}}
\end{split}
\end{equation}
To obtain the PDF at ~$\theta$, we substitute Equation (\ref{eqn:14}) into (\ref{eqn:10}),
\begin{equation}
\label{eqn:17}
f(\theta) = \frac{2\Big(\pi-\arccos(\frac{r_{1}^{2} +r_{2}^{2} - t_{WLAN}^{2}v^{2} }{2r_{1}r_{2}})\Big)}{\pi^{2}}
\end{equation}
Thus, from Equations (\ref{eqn:16}) and (\ref{eqn:17}), we can now obtain the PDF of the traversal time, ~$f(T)$, using Equation (\ref{eqn:13}),
\begin{equation}
\label{eqn:18}
f(T) = \frac{4v^{2}t_{WLAN}\Big(\pi-\arccos(\frac{r_{1}^{2} +r_{2}^{2} - t_{WLAN}^{2}v^{2} }{2r_{1}r_{2}})\Big)}{\pi^{2}\sqrt{4r_{1}^{2}r_{2}^{2} - (r_{1}^{2} +r_{2}^{2} - t_{WLAN}^{2}v^{2})^{2}}}
\end{equation}

\subsection{Handover Probabilities}
\label{sec:4}
To have unnecessary handover and handover failure within satisfactory bounds, it is imperative to find two time threshold values, ~$N$ and ~$M$, which correspond to the values for handover decision for unnecessary handover and handover failure respectively. In order to keep the unnecessary handover and handover failure within bounds the handover will only be initiated if the expected traversal time through the WLAN cell exceeds the corresponding threshold value.

\subsubsection{Probability of Unnecessary Handover}
\label{sec:4a}
This paper attempts to minimize the number of unnecessary handovers. This is achieved by calculating the time threshold value,~$N$, and avoiding handover attempts for which the traversal time through the target network is less than this threshold value. An unnecessary handover is said to occur when the traversing time of an MN in a WLAN cell is smaller than the sum of the handover time into ~$(\tau_A)$ and out of ~$(\tau_D)$ the WLAN coverage area \cite{RefJ3}. We now use the PDF of traversal time obtained in Equation (\ref{eqn:18}) to derive an expression for the CDF of the traversal time, ~$P_u$. This is shown in Equation (\ref{eqn:21}),

\begin{equation}
\label{eqn:19}
P_{u}=\left\{
\begin{array}{c l}
    P_{r}[N<T\leq\tau_{T}], & 0\leq T\leq\frac{(r_{1} +r_{2})}{v},\\
    0, & Otherwise.
\end{array}\right.
\end{equation}

\begin{equation}
\label{eqn:20}
P_{r}[N<T\leq\tau_{T}] = \int^{\tau_{T}}_{N} f(T)\,dt
\end{equation}
Where ~$\tau_{T} = \tau_{A} + \tau_{D}$. The probability of unnecessary handover, ~$P_u$, is expressed as,

\begin{equation}
\label{eqn:21}
\begin{split}
 P_u&=\frac{\Big[2\pi - \arccos(\frac{r_{1}^{2} +r_{2}^{2} - \tau_{T}^{2}v^{2} }{2r_{1}r_{2}})\Big] \arccos\Big(\frac{r_{1}^{2} +r_{2}^{2} - \tau_{T}^{2}v^{2} }{2r_{1}r_{2}}\Big)}{\pi^{2}}  \\
    &-\frac{\Big[2\pi - \arccos(\frac{r_{1}^{2} +r_{2}^{2} - N^{2}v^{2} }{2r_{1}r_{2}})\Big] \arccos\Big(\frac{r_{1}^{2} +r_{2}^{2} - N^{2}v^{2} }{2r_{1}r_{2}}\Big)}{\pi^{2}}
\end{split}
\end{equation}
Let ~$z = \arccos\Big(\frac{r_{1}^{2} +r_{2}^{2} - \tau_{T}^{2}v^{2} }{2r_{1}r_{2}}\Big) $. We obtain the following expression for ~$N$, which is a function of handover latency, velocity, stochastic coverage radius and probability of unnecessary handover.

\begin{equation}
\label{eqn:23}
N = \frac{\sqrt{r_{1}^{2} +r_{2}^{2} - 2r_{1}r_{2}\cos(y)}}{v}
\end{equation}
Where,
\begin{equation}
\label{eqn:22}
y =\pi \pm \sqrt{\pi^{2}(1+P_{u}) -2\pi z + z^{2}}
\end{equation}

We have obtained a new expression for time threshold $N$ for unnecessary handover\footnote{Yan et al.\cite{RefJ2} arrived at a time threshold, ~$t_{WLAN} = \frac{2R}{v}\sin\Big(\arcsin(\frac{v\tau}{2R}-\frac{\pi}{2}P)\Big)$, \\ and Hussain et al.\cite{RefJ3} arrived at, ~$t_{WLAN} = \frac{2Rk}{v\sqrt{1+k^2}}$, where ~$k = \tan\Big[\arctan(\frac{v\tau}{\sqrt{4R^2 - v^2\tau^2}})-\frac{P\pi}{2}\Big]$   }.

\subsubsection{Probability of Handover Failure}
\label{sec:4b}

Handover failure is said to occur if the handover time into ~$(\tau_A)$ the WLAN cell exceeds the overall time spent by the MN in the WLAN coverage area \cite{RefJ3}. A time threshold, ~$M$, is determined and the probability of handover failure is kept within desirable bounds.

\begin{equation}
\label{eqn:24}
P_{f}  =\left\{
\begin{array}{c l}
    P_{r}[M<T\leq\tau_{A}], & 0\leq T\leq\frac{(r_{1} +r_{2})}{v},\\
    0, & Otherwise.
\end{array}\right.
\end{equation}

\begin{equation}
\label{eqn:25}
P_{r}[M<T\leq\tau_{A}] = \int^{\tau_{A}}_{M} f(T)\,dt
\end{equation}
The probability of handover failure, ~$P_f$, is expressed as,
\begin{equation}
\label{eqn:26}
\begin{split}
 P_f&=\frac{\Big[2\pi - \arccos(\frac{r_{1}^{2} +r_{2}^{2} - \tau_{A}^{2}v^{2} }{2r_{1}r_{2}})\Big] \arccos\Big(\frac{r_{1}^{2} +r_{2}^{2} - \tau_{A}^{2}v^{2} }{2r_{1}r_{2}}\Big)}{\pi^{2}}  \\
    &-\frac{\Big[2\pi - \arccos(\frac{r_{1}^{2} +r_{2}^{2} - M^{2}v^{2} }{2r_{1}r_{2}})\Big] \arccos\Big(\frac{r_{1}^{2} +r_{2}^{2} - M^{2}v^{2} }{2r_{1}r_{2}}\Big)}{\pi^{2}}
\end{split}
\end{equation}

Hence, we have also obtained a new expression for time threshold ~$M$, for handover failure control.

\begin{equation}
\label{eqn:28}
M = \frac{\sqrt{r_{1}^{2} +r_{2}^{2} - 2r_{1}r_{2}\cos(q)}}{v}
\end{equation}

Where,
\begin{equation}
\label{eqn:27}
q =\pi \pm \sqrt{\pi^{2}(1+P_{f}) -2\pi z + z^{2}}
\end{equation}

\section{Handover Triggering Condition Estimation}
In this section, we present a handover triggering condition estimation (HTCE) method that attempts to estimate the optimal handover triggering point at which an MN will need to initiate a handover from it's present access network (WLAN) back to the 3G cellular network. HTCE helps to get the best time to initiate a handover in order to avoid connection breakdowns as well as maximize the usage of the preferred network.
\begin{figure*}[h]
\begin{center}
 \includegraphics[width=1\textwidth]{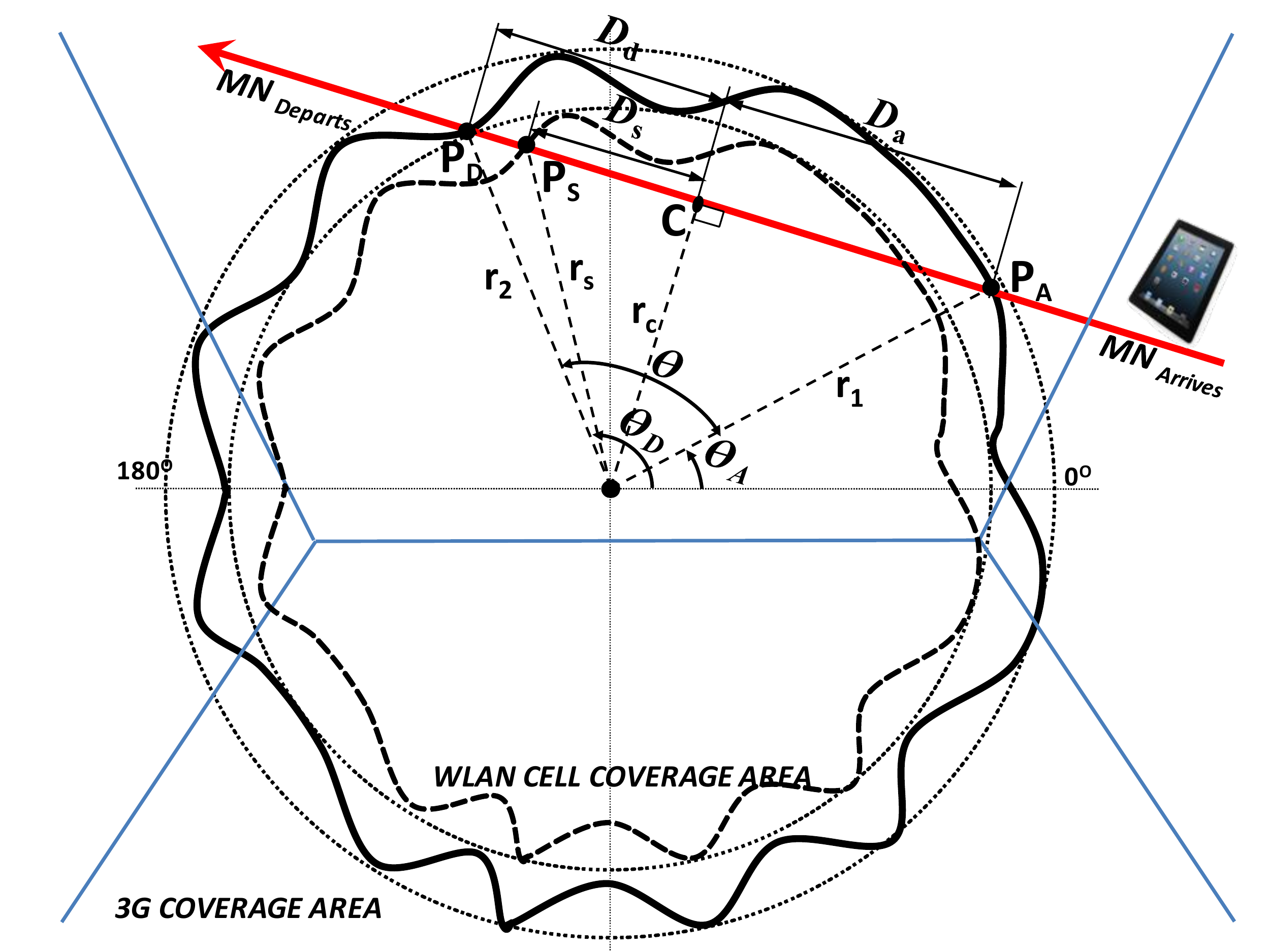}
\caption{Handover triggering condition estimation of an MN in a WLAN cell}
\label{fig:3c}       
\end{center}
\end{figure*}
The path of an MN moving over an area where 3G cellular network service is available and partly covered with a WLAN cell is shown in Fig. \ref{fig:3c}. As the MN approaches the boundary (at the exit point) of the WLAN coverage area, it is necessary for a handover to be triggered in order to bring the probability of connection breakdown to a minimum. The probability of connection breakdown is the ratio of number of connection breakdowns to the total number of handovers. Depending on the requirements set by the system designer, the RSS trigger threshold is usually set to keep the number of connection breakdowns within tolerable bounds. From Fig. \ref{fig:3c}, as the MN approaches the boundary of the WLAN cell, it is best for a handover to be triggered at the point,~$P_{s}$ because if the MN moves beyond this point, the ~$RSS$ from the WLAN drops below a threshold, ~$RSS_{Thresh}$.

The work also assumes that when an MN is in the coverage area of it's present access network, adjacent to a WLAN cell,\footnote{We present an amoebic based model that extends the ideal circle model employed in previous works \cite{RefJ1,RefJ2,RefJ3,RefJ4}} it may enter the boundary area of the WLAN cell at any point, ~$P_A$ and move along the path ~$|P_AP_D|$, making exit from any point, ~$P_D$ on the coverage boundary (as shown in Fig. \ref{fig:3c}).
In order to get the appropriate point,~$P_{s}$ to trigger the handover, we need to determine the radius of the inner amoebic shaped coverage area as shown in Fig. \ref{fig:3c}. This radius varies with time due to the stochastic nature of the wireless medium. Thus, we present an algorithm to determine the value of ~$r_{s}$ and ~$RSS_{Thresh}$.\\
Using the cosine formula,
\begin{equation}
\label{eqn:3.1}
D^2 = r_{1}^{2} +r_{2}^{2} - 2r_{1}r_{2}\cos\theta
\end{equation}
Where ~$D=D_a + D_d$ and ~$C$ is the a point along ~$|P_AP_D|$. Applying Pythagoras theorem, we get,
\begin{equation}
\label{eqn:3.2}
r_c^2 = r_2^2 - D_d^2
\end{equation}
\begin{equation}
\label{eqn:3.3}
D_s^2 = r_s^2 - r_c^2
\end{equation}
Simplifying,
\begin{equation}
\label{eqn:3.4}
D_d = \sqrt{r_{1}^{2} +r_{2}^{2} - 2r_{1}r_{2}\cos\theta} - D_a
\end{equation}
A handover is triggered if the MN leaves the point,~$C$, and RSS from the WLAN drops below or has never been above the threshold,~$RSS_{Thresh}$. Thus, the traversal time,~$t_t$, of the MN in the boundary area of the WLAN cell, i.e. traveling from~$C$ to~$P_D$, is determined as
\begin{equation}
\label{eqn:3.6}
t_t = \frac{D_d}{v}
\end{equation}
By substituting Equation (\ref{eqn:3.4}) in Equation (\ref{eqn:3.6}),~$t_t$ is expressed as
\begin{equation}
\label{eqn:3.7}
t_t = \frac{\sqrt{r_{1}^{2} +r_{2}^{2} - 2r_{1}r_{2}\cos\theta} - D_a}{v}
\end{equation}
The CDF of~$t_t$,~$F(T)$ is expressed as,
\begin{equation}
\label{eqn:3.8}
\begin{split}
 F(T)&=P(T \leq t_t)\\
    &= \left\{
\begin{array}{c l}
    p=P\bigg[T\leq \frac{D_d}{v}\bigg], &   0\leq T\leq \frac{1}{v}\sqrt{r_2^2 -r_s^2},\\
    1, &   Otherwise.
\end{array}\right.
\end{split}
\end{equation}
The probability of~$T\leq \frac{D_d}{v}$,~$p$, is calculated as,
\begin{equation}
\label{eqn:3.9}
\begin{split}
 p&=P\bigg[T\leq \frac{D_d}{v}\bigg]\\
    &= P\Bigg[T\leq \frac{\sqrt{r_{1}^{2} +r_{2}^{2} - 2r_{1}r_{2}\cos\theta} - D_a}{v}\Bigg]\\
    &= P\Bigg[Cos(\theta) \leq \frac{r_1^2 +r_2^2-2D_aTv -T^2v^2 -D_a^2}{2r_1r_2}\Bigg]\\
    &= P\Bigg[0\leq \theta \leq \arccos\Bigg(\frac{r_1^2 +r_2^2-2D_aTv -T^2v^2 -D_a^2}{2r_1r_2}\Bigg)\Bigg]\\
    &= P\Bigg[\pi - \arccos\Bigg(\frac{r_1^2 +r_2^2-2D_aTv -T^2v^2 -D_a^2}{2r_1r_2}\Bigg) \leq \theta \leq \pi\Bigg]
\end{split}
\end{equation}
Integrating~$f(\theta)$ from Equation (\ref{eqn:10}), we get the CDF which is within the range stated in Equation (\ref{eqn:3.9})
\begin{equation}
\label{eqn:3.10}
\begin{split}
p&=\int_{0}^{\arccos\Big(\frac{r_1^2 +r_2^2-2D_aTv -T^2v^2 -D_a^2}{2r_1r_2}\Big)} f(\theta)d\theta + \int_{\pi - \arccos\Big(\frac{r_1^2 +r_2^2-2D_aTv -T^2v^2 -D_a^2}{2r_1r_2}\Big)}^{\pi} f(\theta)d\theta\\
&=1-\pi +\frac{2}{\pi}\arccos\Bigg(\frac{r_1^2 +r_2^2-2D_aTv -T^2v^2 -D_a^2}{2r_1r_2}\Bigg)
\end{split}
\end{equation}
A connection breakdown is said to occur if the traversal time in the boundary area is less than the handover delay from the WLAN to the cellular network,~$\tau_{D}$. The connection breakdown probability, ~$P_{Break}$ is calculated as,
\begin{equation}
\label{eqn:3.11}
P_{Break} = \left\{
\begin{array}{c l}
    1, &   r_2<r_s,\\
    0, &   \tau_{D} < \frac{r_2 - r_s}{v},\\
    1-\pi +\frac{2}{\pi}\arccos\bigg(\frac{r_1^2 +r_2^2-2D_a\tau_{D}v -\tau_{D}^2v^2 -D_a^2}{2r_1r_2}\bigg), &   Otherwise.
\end{array}\right.
\end{equation}
Thus, we can estimate the value of~$r_s$ for any given breakdown probability,~$P_{Break}$, within ~$0<P_{Break}<1$,
\begin{equation}
\label{eqn:3.12}
r_s= \sqrt{\bigg[r_1\Psi - \sqrt{D_a^2 - r_1^2 + 2D_a\tau_{D}v +\tau_{D}^2v^2+ r_1^2\Psi^2}\bigg]^2 - C_{a}}
\end{equation}
Where~$\Psi = \cos\Big[\frac{\pi}{2}(P_{Break} + \pi - 1)\Big]$ and~$C_{a}$ is the channel adjustment parameter.\\
RSS measurements are widely used to estimate distance due to the fact that they require no additional hardware, however, shadowing degrades the accuracy of estimation significantly \cite{RefJ11}. This study integrates the effect of shadow fading in estimating the handover triggering condition of an MN traversing a WLAN to a 3G cellular coverage area. Finally, by applying the log-distance path loss model \cite{RefB2},~$RSS_{Thresh}$ is
expressed as
\begin{equation}
\label{eqn:3.12b}
RSS_{Thresh}= P(transmit) - PL(d_0) - 10\beta\log_{10}(r_s/d_0) + R_{\sigma}
\end{equation}
Where~$P(transmit)$ is the transmit power of the WLAN AP in dBm,~$d_0$ is the distance between the AP and a reference point,~$PL(d_0)$ is the path loss at the reference point in dB,~$\beta$ is the path loss exponent, and~$R_{\sigma}$ is a Gaussian distributed random variable with a mean of zero and a standard deviation,~$\sigma$ in dB.
\subsection{HTCE Optimization}

A handover should be initiated when the RSS level of the serving access network drops below the handover threshold,~$RSS_{Thresh}$. Due to the limiting effect of estimating triggering conditions, we present an adaptive handover threshold that considers the effect of shadowing to ensure seamless handover as an MN moves from a WLAN to a 3G coverage area. Seamless handover is a function of the handover threshold,~$RSS_{Thresh}$, thus, the~$RSS_{Thresh}$ value must be precisely determined to avoid connection breakdown and loss of data packets at the boundary area of the serving access network.

From Fig. \ref{fig:3c}, we assume that the boundary distance,~$d_B = D_d - D_s$ and~$t_s$ is the instant at~$P_S$ when the RSS from the serving WLAN AP drops below the handover threshold,~$RSS(t_s)\leq RSS_{Thresh}$. When this occurs, HTS is triggered and a handover may occur. Let~$T_B$ be the duration of the MN at the boundary area (ie. from the point~$P_S$ to~$P_D$). The distance between the critical point,~$P_S$ and boundary point,~$P_D$ is called the boundary traversal distance,~$d_B$. The adaptive handover threshold is now a function of the boundary traversal distance,~$d_B$. We have
\begin{equation}
\label{eqn:3.13}
T_B = \frac{d_B}{v} \triangleq \frac{\chi r_2}{v}
\end{equation}
Where~$\chi \in [0,1]$,~$r_2$ is the estimated random cell radius and~$v$ is the velocity of the MN along the boundary region.
We also assume the packet delay from the instant when the current serving AP sends a packet to when the packet arrives at the MN is given as~$\delta$. In order to have seamless connectivity with no loss of packets, the last packets sent by the WLAN AP must arrive at the MN before the MN moves out of the boundary region. To achieve seamless handover, Equation (\ref{eqn:3.14}) must be satisfied.
\begin{equation}
\label{eqn:3.14}
T_B \geq \tau_B + \delta
\end{equation}

Where~$\tau_B$ is the handover latency from the instant~$t_s$ to the instant when the WLAN AP (current serving access network) receives a handover notification from the MN. Immediately the WLAN AP receives the routing information update from the MN, data packets will be routed to the MN through a new established path. From Equation (\ref{eqn:3.13}) and (\ref{eqn:3.14}), we have,
\begin{equation}
\label{eqn:3.15}
\chi r_2 \geq (\tau_B + \delta)v
\end{equation}
By implication, handover should be initiated at distance~$(1-\chi)r_2$ from the WLAN AP. Thus, from the signal propagation model in Equation (\ref{eqn:3.12b}), we can express the random RSS at~$(1-\chi)r_2$,~$RSS_{Thresh}$ as
\begin{equation}
\begin{split}
\label{eqn:3.16}
RSS_{Thresh}&= P(transmit) + Gains - Losses + R_{\sigma}\\
            &=F_1 - F_2\log_{10}([1-\chi]r_2)
\end{split}
\end{equation}
Where~$F_1 = P(transmit) + Gains + R_{\sigma}$,~$F_2=10\beta$ and~$\beta$ is the path loss exponent. It is noteworthy that the random variable,~$R_{\sigma}$ is a zero mean stationary Gaussian random process modeling shadow fading in the channel. The WLAN cell radius can be estimated based on RSS measurements, the velocity of the MN can also be estimated, the handover delay to a given access network is usually a known parameter. The packet delay,~$\delta$ can also be estimated based on the Round Trip Time (RTT) value. From Equation (\ref{eqn:3.15}) and (\ref{eqn:3.16}), we have,
\begin{equation}
\begin{split}
\label{eqn:3.17}
RSS_{Thresh}&= F_1 - F_2\log_{10}\Big(1-\frac{(\tau_B + \delta)v}{r_2}\Big) - F_2\log_{10}(r_2)\\
            &= RSS_{B} - F_2\log_{10}\Big(1-\frac{(\tau_B + \delta)v}{r_2}\Big)
\end{split}
\end{equation}
Where~$RSS_{B}$ is the RSS at the border of the WLAN coverage cell.
From Equation (\ref{eqn:3.15}), we see that packet loss occur~$\iff\tau_B + \delta - T_B > 0$ and we now consider a fixed handover threshold,~$RSS_{Thresh}^{fixed}$, expressed as,
\begin{equation}
\label{eqn:3.18}
RSS_{Thresh}^{fixed}= RSS_{B} - F_2\log_{10}\Big(1-\frac{vT_B}{r_2}\Big)
\end{equation}
Assuming data packets are transmitted by the WLAN AP at constant data rate,~$\Re$, the total number of packets loss,~$P_{Loss}$ is given as,
\begin{equation}
\label{eqn:3.19}
P_{Loss}= (\tau_B + \delta - T_B)\Re
\end{equation}
From Equation (\ref{eqn:3.18}), we get an expression for~$T_B$ and for~$(\tau_B + \delta)$ from Equation (\ref{eqn:3.17}). We then substitute them into Equation (\ref{eqn:3.19}), thus, we get an expression for the number of loss packets due to fixed handover threshold,
\begin{equation}
\label{eqn:3.20}
P_{Loss}= \bigg(10^{\frac{RSS_{B} - RSS_{Thresh}^{fixed}}{F_2}} - 10^{\frac{RSS_{B} - RSS_{Thresh}^{adaptive}}{F_2}}\bigg)\frac{r_{2} \Re}{v}
\end{equation}
Where~$RSS_{Thresh}^{adaptive}$ is the adaptive handover threshold and~$\Delta RSS_{Thresh}$ is the difference between the RSS at the border area and the RSS threshold value.
\section{Handover Target Selection}
\begin{figure*}[h]
\begin{center}
 \includegraphics[width=1.1\textwidth]{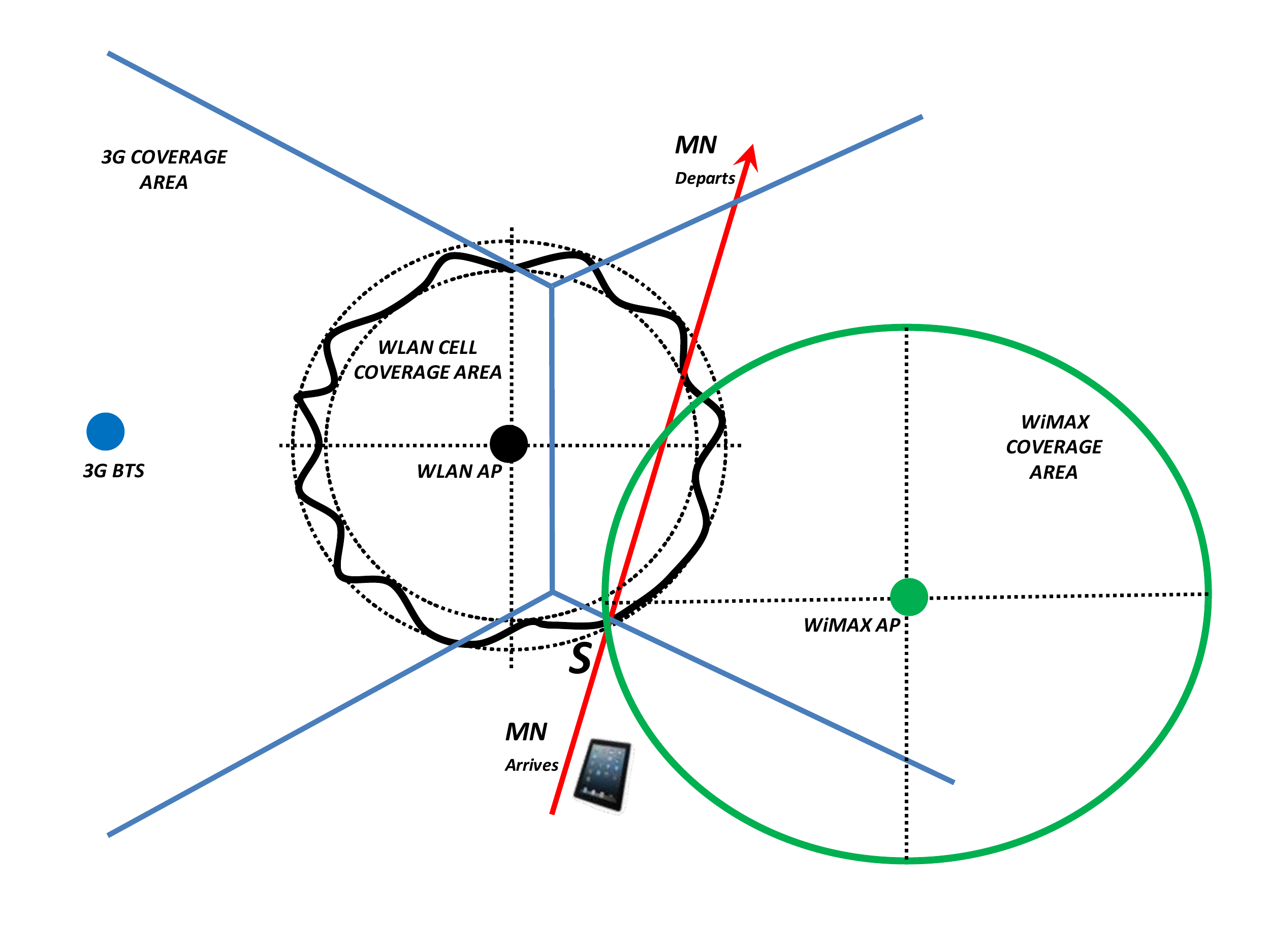}
\caption{A mobile node making a Target Selection}
\label{fig:3b}       
\end{center}
\end{figure*}

In this section, two case studies were considered for implementing the GRA procedure. Fig. \ref{fig:3b} shows an MN traversing a heterogeneous environment. At point~$S$, the MN node is in the range of different available access networks. The selection of the optimal access network to handover to becomes a big challenge. This problem can only be solved using a multi-criteria selection algorithm. Thus, the study presents the implementation of the GRA procedure using practical examples.
\subsection{Grey Relational Analysis Approach}
The proposed GRA procedure is applied to provide a basis for a faster and accurate network selection decision making process. Computational simulations were carried out using MATLAB. Both examples deal with the selection of an optimal access network in a heterogeneous environment. The data used are experimental and may vary from real-world scenarios. They are only used to illustrate the GRA procedure in the presence of multiple access networks.

\subsubsection{Case Study One}
The first example considers three available access networks (WLAN, WiMAX and 3G) and five performance attributes (Cost, Delay, Data rates, Dwell Time\footnote{A novel performance criterion used in the GRA approach and can be obtained using the HNE model}, QoS). Out of the five\footnote{We apply Equation (\ref{eqn:2d1}) and (\ref{eqn:2d2}) to solve advantageous and non-advantageous performance attributes respectively}, cost and delay are non-advantageous performance attributes, while the others are advantageous. Table \ref{tab:3a} shows the network selection decision matrix with various parameters of different SI units.
\newpage
\begin{table}[H]
\begin{tcolorbox}[tab2,tabularx={X||Y|Y|Y|Y|Y}]

\begin{flushleft}
\textbf{Access Network}
\end{flushleft}

 &
\begin{center}
\textbf{Cost ~$(\$/MB)$}
\end{center}
 &
 \begin{center}
 \textbf{Delay ~$(ms)$}
 \end{center}
 &
 \begin{center}
 \textbf{Data Rate ~$(Mbps)$}
 \end{center}
 &
 \begin{center}
 \textbf{Dwell Time ~$(min)$}
 \end{center}
 &
 \begin{center}
 \textbf{QoS ~$(1-5)$}
 \end{center}
     \\\hline\hline
WLAN   & 0.20 & 132.00 &  12.00 &  5.00 & 4.00 \\\hline
WiMAX & 0.40 & 140.00 &  8.00 &  8.00 & 3.00 \\\hline
3G  & 1.30 & 162.00 &  2.00 &  15.00 & 2.00 \\
\end{tcolorbox}
\caption{Network Selection Decision Matrix for Case Study 1}
\label{tab:3a}
\end{table}
In Table \ref{tab:3b}, the study assigned optimal reference value of 100\%,~$X_0 = [1, 1, 1, 1, 1]$ for all five performance attributes.
\begin{table}[H]
\begin{tcolorbox}[tab2,tabularx={X||Y|Y|Y|Y|Y}]
\begin{flushleft}
\textbf{Access Network}
\end{flushleft}

 &
\begin{center}
\textbf{Cost ~$(\$/MB)$}
\end{center}
 &
 \begin{center}
 \textbf{Delay ~$(ms)$}
 \end{center}
 &
 \begin{center}
 \textbf{Data Rate ~$(Mbps)$}
 \end{center}
 &
 \begin{center}
 \textbf{Dwell Time ~$(min)$}
 \end{center}
 &
 \begin{center}
 \textbf{QoS ~$(1-5)$}
 \end{center}
     \\\hline\hline
REF.~$X_{0}$ & 1.0000 & 1.0000 &  1.0000 &  1.0000 & 1.0000 \\\hline
WLAN & 1.0000 & 1.0000 &  1.0000 &  0.0000 & 1.0000 \\\hline
WiMAX & 0.5775 & 0.9091 &  0.6000 &  0.3636 & 0.6667 \\\hline
3G  & 0.0000 & 0.0000 &  0.0000 &  1.0000 & 0.0000 \\
\end{tcolorbox}
\caption{Normalized Decision Matrix for Case Study 1}
\label{tab:3b}
\end{table}
The study used a distinguishing coefficient($\zeta$) of 0.5 to calculate the GRC. Using Equation \ref{eqn:2d4}, Table \ref{tab:3c} shows the obtained results.
\newpage
\begin{table}[H]
\begin{tcolorbox}[tab2,tabularx={X||Y|Y|Y|Y|Y}]
\begin{flushleft}
\textbf{Access Network}
\end{flushleft}

 &
\begin{center}
\textbf{Cost ~$(\$/MB)$}
\end{center}
 &
 \begin{center}
 \textbf{Delay ~$(ms)$}
 \end{center}
 &
 \begin{center}
 \textbf{Data Rate ~$(Mbps)$}
 \end{center}
 &
 \begin{center}
 \textbf{Dwell Time ~$(min)$}
 \end{center}
 &
 \begin{center}
 \textbf{QoS ~$(1-5)$}
 \end{center}
     \\\hline\hline
WLAN   & 1.0000 & 1.0000 &  1.0000 &  0.3333 & 1.0000 \\\hline
WiMAX & 0.5420 & 0.8462  &  0.5556 &  0.4400 & 0.6000 \\\hline
3G  & 0.3333 & 0.3333 &  0.3333 &  1.0000 & 0.3333 \\
\end{tcolorbox}
\caption{Grey Relational Coefficient Matrix for Case Study 1}
\label{tab:3c}
\end{table}
$\Gamma(X_{0}, X_{i})$ calculation was done using Equation \ref{eqn:2d5} and the result is shown in Table \ref{tab:3d}.
\begin{table}[H]
\begin{tcolorbox}[tab2,tabularx={X||Y|Y}]
\begin{flushleft}
\textbf{Access Network}
\end{flushleft}

 &
\begin{center}
\textbf{~$\Gamma(X_{0}, X_{i})$}
\end{center}
 &
 \begin{center}
 \textbf{GRA Rank}
 \end{center}
      \\\hline\hline
WLAN   & 0.8667 & ~$1^{st}$ \\\hline
WiMAX & 0.5967 & ~$2^{nd}$  \\\hline
3G  & 0.4667 & ~$3^{rd}$  \\
\end{tcolorbox}
\caption{GRA Ranking for Case Study 1}
\label{tab:3d}
\end{table}

\subsubsection{Case Study Two}
To further elaborate on the speed and efficiency of the GRA approach, the study considers another case study with five available networks (i.e. two WLAN, two WiMAX and a 3G network are available) and six performance attributes\footnote{We apply Equation (\ref{eqn:2d1}) and (\ref{eqn:2d2}) to solve advantageous and non-advantageous performance attributes respectively} (i.e. cost, delay, data rate, dwell time\footnote{A novel performance criterion used in the GRA approach and can be obtained using the HNE model}, observable Qos and RSS values) as seen in Table \ref{tab:3e}. Out of all the performance attributes considered, only cost and delay are non-advantageous (with the lower the value, the better). GRA steps is implemented on the data with results shown in Table \ref{tab:3f}, \ref{tab:3g}, \ref{tab:3h}.
\newpage
\begin{table}[h]
\begin{tcolorbox}[tab2,tabularx={X||Y|Y|Y|Y|Y|Y}]
\begin{flushleft}
\textbf{Access Network}
\end{flushleft}

 &
\begin{center}
\textbf{Cost ~$(\$/MB)$}
\end{center}
 &
 \begin{center}
 \textbf{Delay ~$(ms)$}
 \end{center}
 &
 \begin{center}
 \textbf{Data Rate ~$(Mbps)$}
 \end{center}
 &
 \begin{center}
 \textbf{Dwell Time ~$(min)$}
 \end{center}
 &
 \begin{center}
 \textbf{QoS ~$(1-5)$}
 \end{center}
 &
 \begin{center}
 \textbf{RSS ~$(dBm)$}
 \end{center}
     \\\hline\hline
WLAN1   & 0.20 & 130.00 &  8.00 &  2.00 & 5.00 & -98.00 \\\hline
WLAN2   & 0.20 & 138.00 &  6.00 &  4.00 & 5.00 & -90.00 \\\hline
WiMAX1 & 0.40 & 132.00 &  10.00 &  13.00 & 4.00 & -72.00 \\\hline
WiMAX2 & 0.40 & 140.00 &  8.00 &  10.00 & 4.00 & -85.00 \\\hline
3G  & 1.62 & 160.00 &  2.00 &  14.00 & 2.00 & -101.00 \\
\end{tcolorbox}
\caption{Network Selection Decision Matrix for Case Study 2}
\label{tab:3e}
\end{table}
In Table \ref{tab:3f}, the study set optimal reference value of 100\%,~$X_0 = [1, 1, 1, 1, 1, 1]$ for all six performance attributes.
\begin{table}[h]
\begin{tcolorbox}[tab2,tabularx={X||Y|Y|Y|Y|Y|Y}]
\begin{flushleft}
\textbf{Access Network}
\end{flushleft}

 &
\begin{center}
\textbf{Cost ~$(\$/MB)$}
\end{center}
 &
 \begin{center}
 \textbf{Delay ~$(ms)$}
 \end{center}
 &
 \begin{center}
 \textbf{Data Rate ~$(Mbps)$}
 \end{center}
 &
 \begin{center}
 \textbf{Dwell Time ~$(min)$}
 \end{center}
 &
 \begin{center}
 \textbf{QoS ~$(1-5)$}
 \end{center}
 &
 \begin{center}
 \textbf{RSS ~$(dBm)$}
 \end{center}
     \\\hline\hline
REF.~$X_{0}$    & 1.0000 & 1.0000 &  1.0000 &  1.0000 & 1.0000 & 1.0000 \\\hline
WLAN1   & 1.0000 & 1.0000 &  0.7500 &  0.0000 & 1.0000 & 0.1034 \\\hline
WLAN2   & 1.0000 & 0.7333 &  0.5000 &  0.1667 & 1.0000 & 0.3793 \\\hline
WiMAX1 & 0.8592 & 0.9333 &  1.0000 &  0.9167 & 0.6667 & 1.0000 \\\hline
WiMAX2 & 0.8592 & 0.6667 &  0.7500 &  0.6667 & 0.6667 & 0.5517 \\\hline
3G  & 0.0000 & 0.0000 &  0.0000 &  1.0000 & 0.0000 & 0.0000 \\
\end{tcolorbox}
\caption{Normalized Decision Matrix for Case Study 2}
\label{tab:3f}
\end{table}
\newpage
The study also used a distinguishing coefficient (($\zeta$)) of 0.5 to calculate the grey relational coefficient. Table \ref{tab:3g} shows the results obtained.
\begin{table}[h]
\begin{tcolorbox}[tab2,tabularx={X||Y|Y|Y|Y|Y|Y}]
\begin{flushleft}
\textbf{Access Network}
\end{flushleft}

 &
\begin{center}
\textbf{Cost ~$(\$/MB)$}
\end{center}
 &
 \begin{center}
 \textbf{Delay ~$(ms)$}
 \end{center}
 &
 \begin{center}
 \textbf{Data Rate ~$(Mbps)$}
 \end{center}
 &
 \begin{center}
 \textbf{Dwell Time ~$(min)$}
 \end{center}
 &
 \begin{center}
 \textbf{QoS ~$(1-5)$}
 \end{center}
 &
 \begin{center}
 \textbf{RSS ~$(dBm)$}
 \end{center}
     \\\hline\hline
WLAN1   & 1.0000 & 1.0000 &  0.6667 &  0.3333 & 1.0000 & 0.3580 \\\hline
WLAN2   & 1.0000 & 0.6522 &  0.5000 &  0.3750 & 1.0000 & 0.4462 \\\hline
WiMAX1 & 0.7802 & 0.8824 &  1.0000  &  0.8571 & 0.6000 & 1.0000 \\\hline
WiMAX2 & 0.7802 & 0.6000 &  0.6667 &  0.6000 & 0.6000 & 0.5273 \\\hline
3G  & 0.3333 & 0.3333 &  0.3333 &  1.0000 & 0.3333 & 0.3333 \\
\end{tcolorbox}
\caption{Grey Relational Coefficient Matrix for Case Study 2}
\label{tab:3g}
\end{table}
The calculation for~$\Gamma(X_{0}, X_{i})$ was done using Equation \ref{eqn:2d5} and the result is shown in Table \ref{tab:3h}.
\begin{table}[H]
\begin{tcolorbox}[tab2,tabularx={X||Y|Y}]
\begin{flushleft}
\textbf{Access Network}
\end{flushleft}

 &
\begin{center}
\textbf{~$\Gamma(X_{0}, X_{i})$}
\end{center}
 &
 \begin{center}
 \textbf{GRA Rank}
 \end{center}

     \\\hline\hline
WLAN1   & 0.7263 & ~$2^{nd}$ \\\hline
WLAN2   & 0.6622 & ~$3^{rd}$ \\\hline
WiMAX1 & 0.8533 & ~$1^{st}$  \\\hline
WiMAX2 & 0.6290 & ~$4^{th}$  \\\hline
3G  & 0.4444 & ~$5^{th}$  \\
\end{tcolorbox}
\caption{GRA Ranking for Case Study 2}
\label{tab:3h}
\end{table}

\section{Summary}
In this chapter, extensive geometric and probability analysis was used to determine suitable handover thresholds for HNE and HTCE. The HNE method calculates threshold values based on various network parameters which include the random varying cell radius, handover latency, the traverse angle,~$\theta$, the velocity of the MN. The HTCE method calculates the RSS threshold for triggering a handover based on the random varying cell radius, handover latency, velocity of the MN and connection breakdown tolerance.  In the proposed HTS approach, the GRA algorithm was applied on two different case studies and used to select an optimal access network to perform a handover based on certain performance criteria.

%% file: chap4.tex
\chapter{Results and Discussion}
\label{chp:4}
\newpage
\section{Introduction}
{\Huge $\mathbb{T}$}his chapter presents simulation results of the proposed scheme. Results for HNE, HTCE and HTS were obtained using MATLAB.
\section{Results for HNE}
\begin{figure*}[h]
\begin{center}
 \includegraphics[scale=0.8]{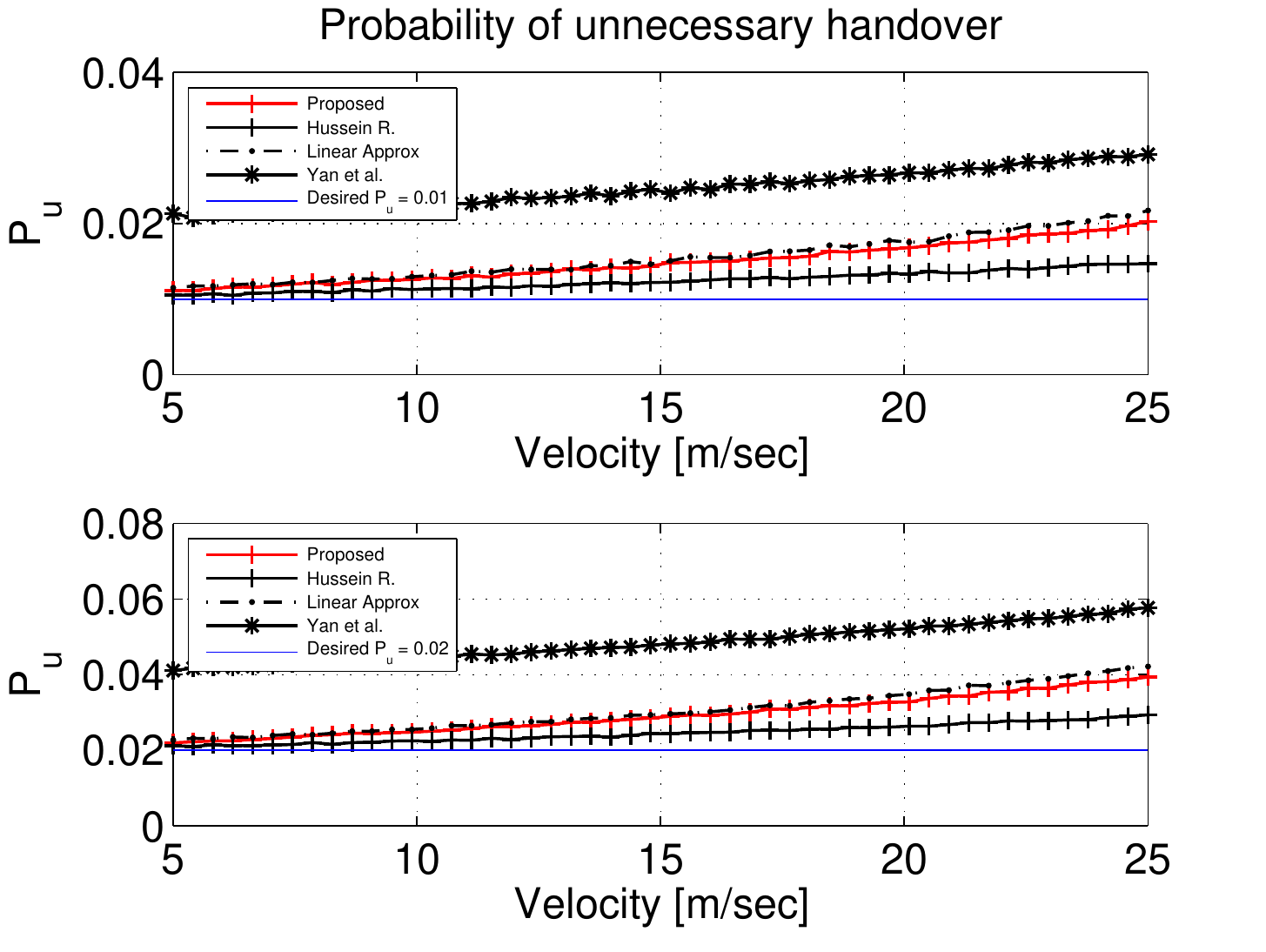}
 \caption{Plot of Probability of Unnecessary Handover vs Velocity of MN.}
\label{fig:4a}       
\vspace{0.6cm}
 \includegraphics[scale=0.8]{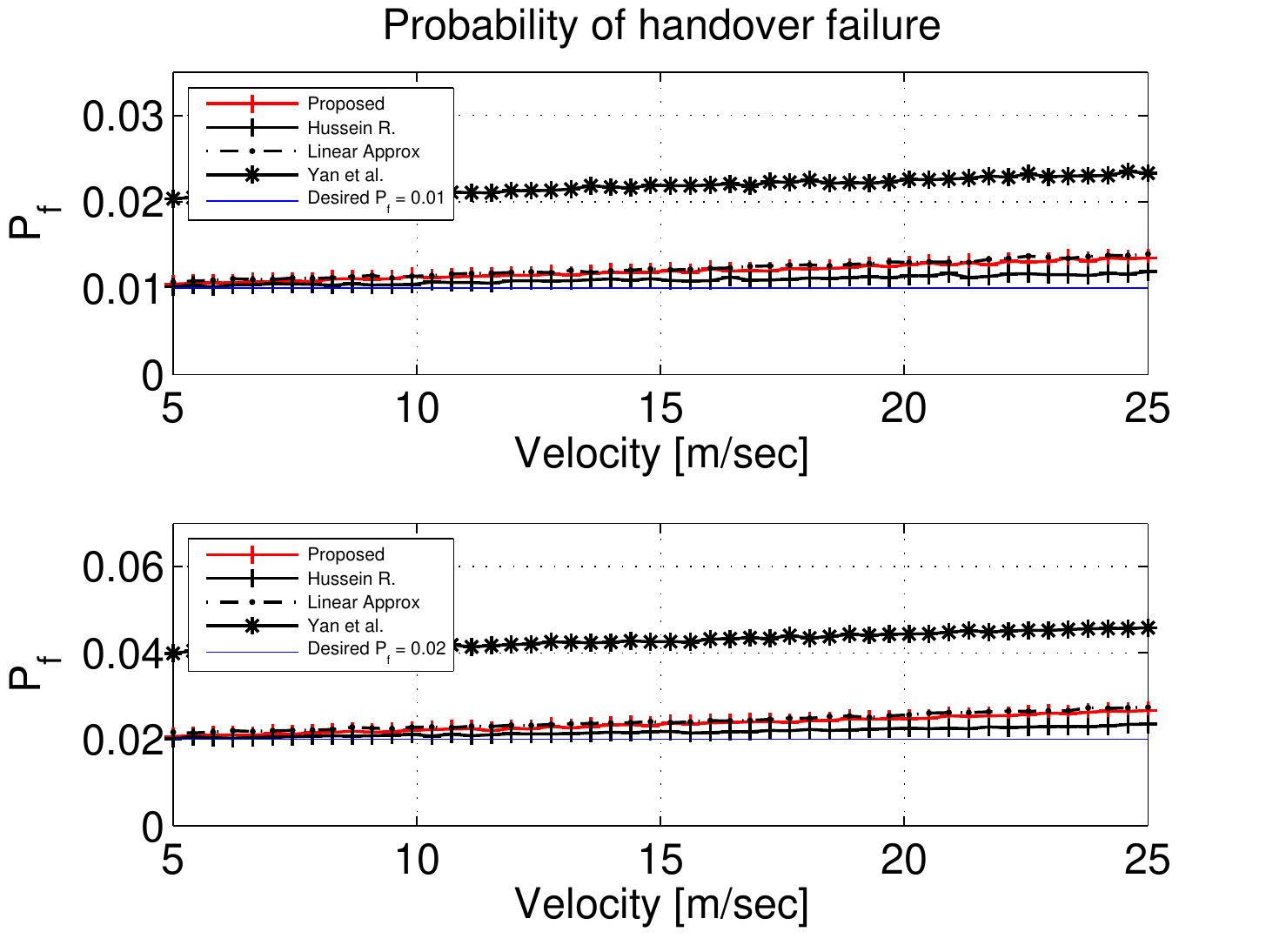}
\caption{Plot of Probability of Handover Failure vs Velocity of MN.}
\label{fig:4b}       
\end{center}
\end{figure*}

To evaluate the proposed HNE model, the study performed Monte-Carlo simulations with about 10 million iterations using MATLAB in order to ensure accurate performance. For validation purpose, we also simulated the performance of the proposed model alongside that of other state-of-the-art time prediction vertical handover schemes. Fig. \ref{fig:4a} and \ref{fig:4b} show the plots of Probability of Unnecessary Handover,~$P_u$ and Probability of Handover Failure,~$P_f$ against the Velocity of MN,~$v$. The threshold values~$(M$ and ~$N)$ obtained in Equations (\ref{eqn:23} and \ref{eqn:28}), the transversal angle,~$\theta$ within ~$[0,\pi]$ bound, dwell time,~$T$ obtained in Equations (\ref{eqn:12}) were used in the experiments. From the graphs, we observe that as the velocity of the MN increases, the probability of unnecessary handover and handover failure increases and deviates from the designed level. This implies that speed has an impact on the prediction of threshold values, which are obtained using this probabilistic model. This deviation in probabilities of handover failure and unnecessary handover are in compliance with the existing works~\cite{RefJ1,RefJ2,RefJ3,RefJ4}.

Results show that the proposed model performs closely to the Linear approximation method employed in \cite{RefJ1} which considered a circular coverage cell. Our work out-performed results of Yan \emph{et al.} \cite{RefJ2}, which considered a ~$[0,2\pi]$ bound, but under-performed when compared to the work of Hussain \emph{et al.} \cite{RefJ3}, which also considered the angle of arrival and departure to lie between ~$[0,\pi]$. However, this model\cite{RefJ3} is not only unrealistic but also impractical as it requires precise information (on the tangential angle of arrival of the MN which is uniformly distributed between ~$[0,\frac{\pi}{2}]$) from the system.\clearpage Despite the performance limitations, our work considered the effect of slow fading and presents a realistic depiction of the WLAN coverage area with the view that the wireless environment is a stochastic one with numerous uncertainties, the proposed model gives a more accurate prediction than previous works in literature.
\section{Results for HTCE}
\begin{figure*}
\begin{center}
 \includegraphics[scale=0.8]{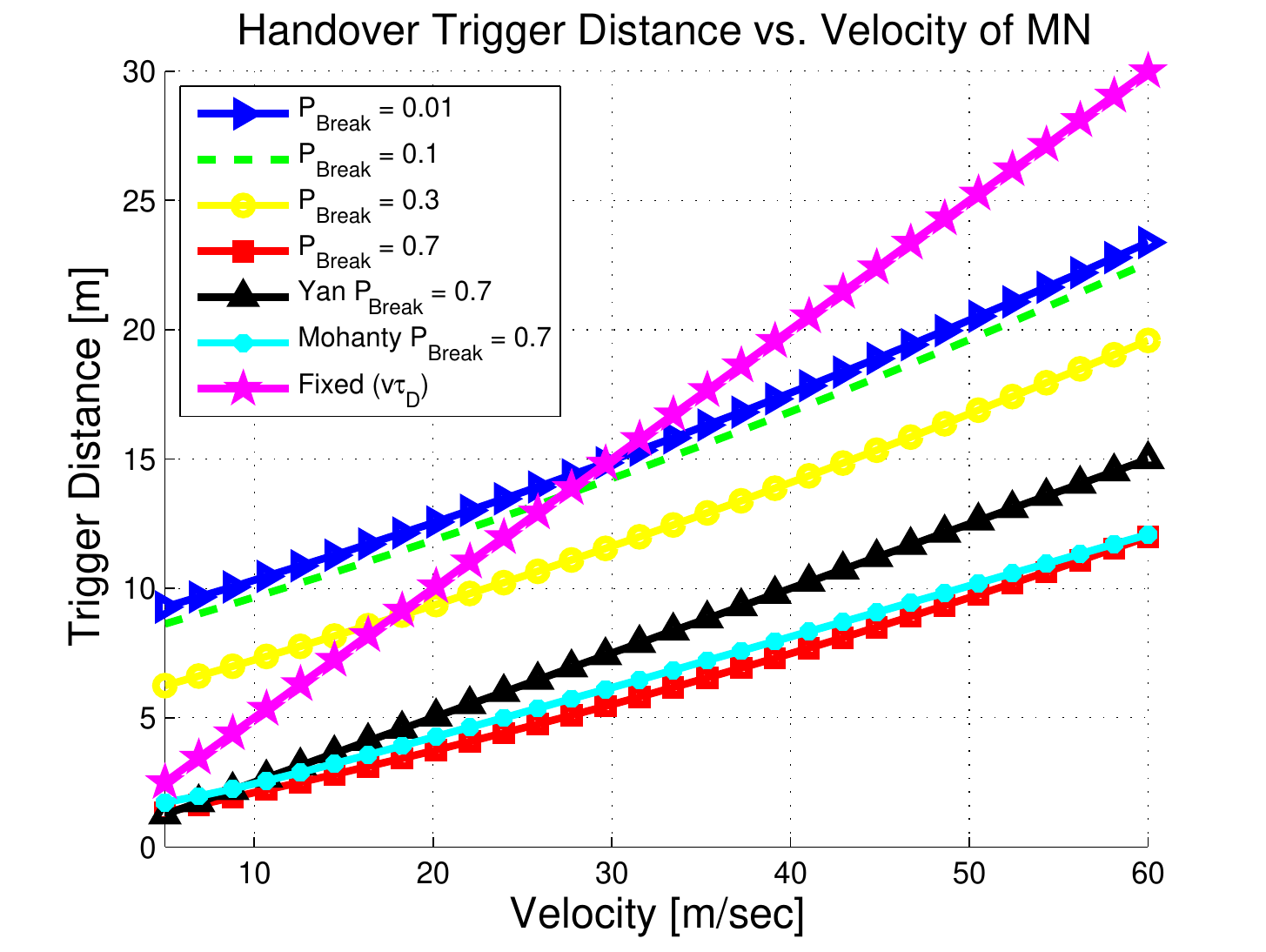}
 \caption{Plot of Handover Trigger Distance vs Velocity of MN.}
\label{fig:4c}       
\end{center}
\end{figure*}

Fig. \ref{fig:4c} shows the handover triggering distance for an ideal case and for various breakdown tolerance values of HTCE. The HTCE algorithm with the lowest breakdown tolerance (~$P_{Break}$ = 0.02) exhibits a similar behavior with other~$P_{Break}$ values, however, it initiates the handover much earlier (At distance~$\approx10m$ with a speed of~$5m/s$) to avoid possible risk of connection breakdown. With higher breakdown tolerance (for~$P_{Break}$ values of 0.3 and 0.7), HTCE takes higher risk of possible connection breakdown in order to increase the time spent in the WLAN coverage area. At~$P_{Break}$ value of 0.7, HTCE delays handover beyond the ideal case. Depending on the decision of the system designer, there should be a trade-off between the number connection breakdowns and the time spent in the WLAN cell. The slope of the proposed HTCE model gives a better response than the ideal case, since it triggers handover at reasonable distances despite high velocity of the MN. These results also demonstrate the flexibility of the proposed HTCE model.

\begin{figure*}
\begin{center}
 \includegraphics[scale=0.8]{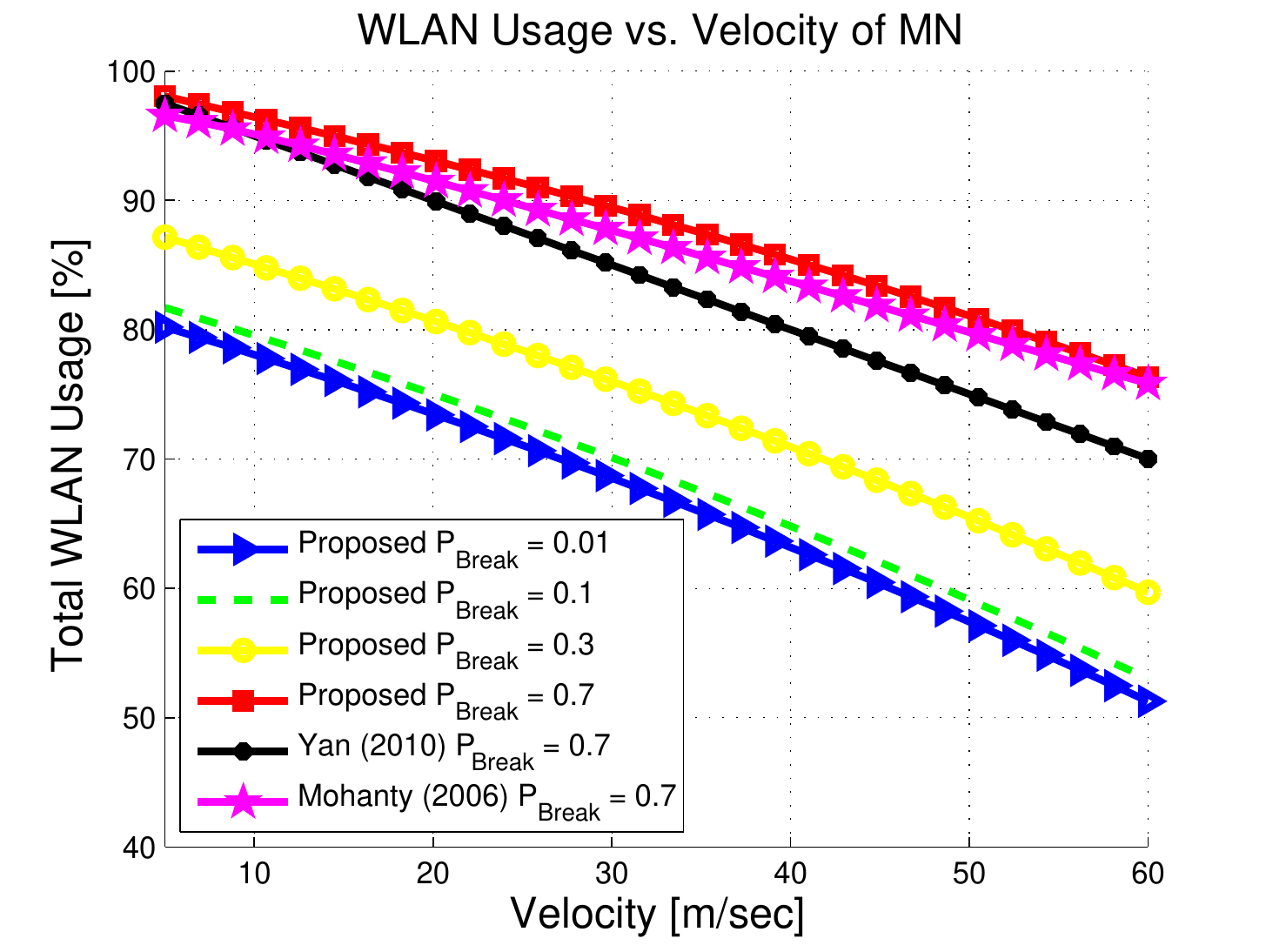}
 \caption{Plot of WLAN Usage vs Velocity of MN.}
\label{fig:4d}       
\end{center}
\end{figure*}

\begin{figure*}[h]
\begin{center}
 \includegraphics[scale=0.8]{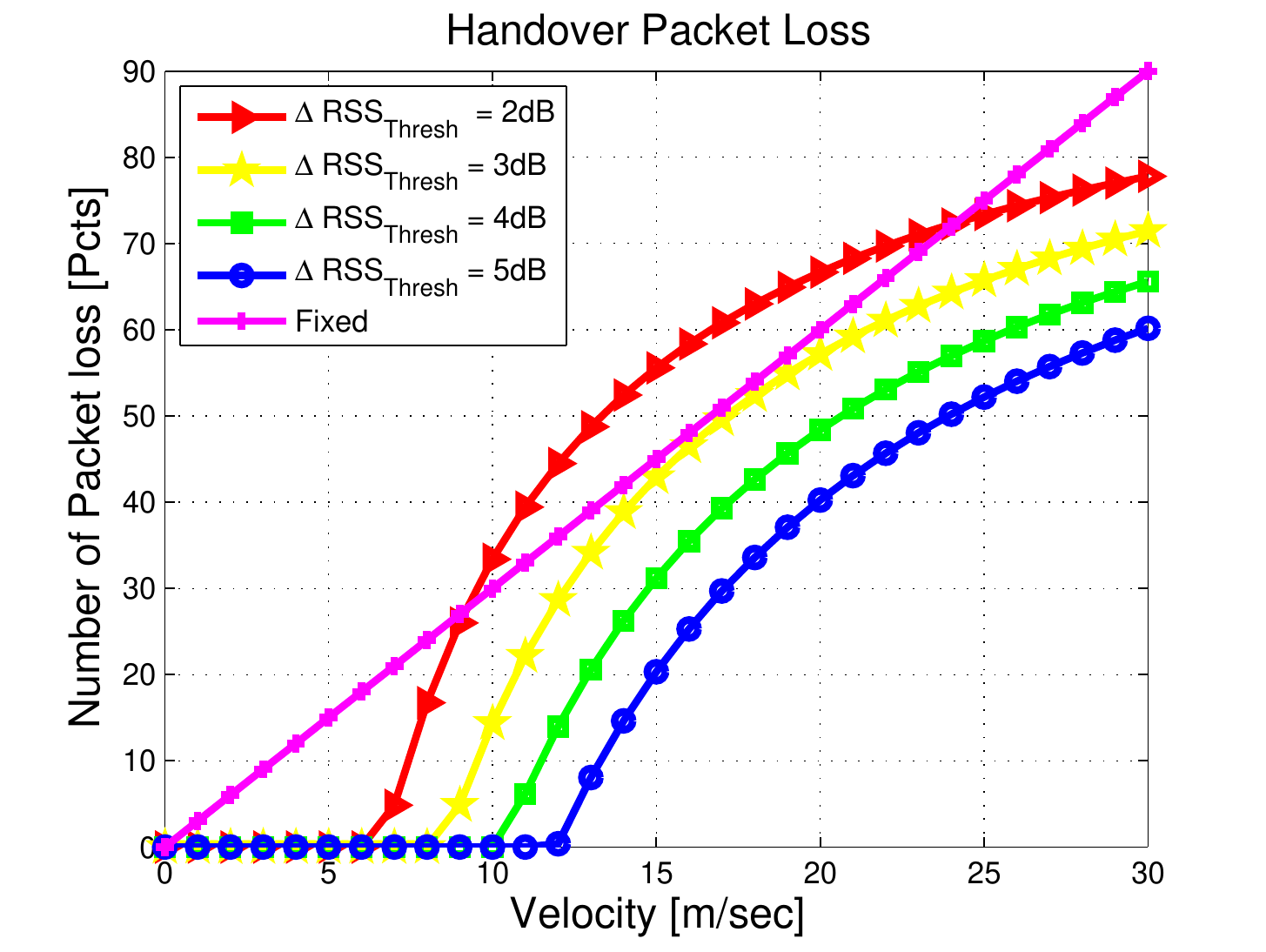}
 \caption{Plot of Packet Loss vs Velocity of MN.}
\label{fig:4e1}       
\end{center}
\end{figure*}

As seen in Fig. \ref{fig:4d}, the total WLAN usage for HTCE increases as higher values of~$P_{Break}$ are applied. Depending on the application being used by the MN, it is possible to adjust the value of~$P_{Break}$ to either maximizing WLAN usage or minimizing the probability of connection breakdown. At~$v=10m/sec$, if the~$P_{Break}$ is set at 0.02, the MN utilizes the WLAN cell for~$\approx 77\%$, while at~$P_{Break}$=0.7, the WLAN cell is utilized for~$\approx 96\%$. For breakdown-sensitive applications such as voice, video calls and networked games,~$P_{Break}$ could be assigned a small value to maintain service continuity, while for breakdown-insensitive applications such as data,~$P_{Break}$ could be assigned a higher value to
maximize WLAN usage (for the purpose of low cost and high bandwidth facility in the WLAN cell).
Fig. \ref{fig:4e1} shows the plot of Packets loss against the velocity of the MN for various fixed threshold values and the results were obtained using the following parameters,~$r_2 = 50 + \aleph[0,5]m$,~$\Re=60packets/sec$ and~$\beta=3$.
From Fig. \ref{fig:4e1}, it can be observed that if~$RSS_{Thresh}^{fixed}<RSS_{Thresh}^{adaptive}$, loss of packet will occur. If~$RSS_{Thresh}^{fixed}$ is assigned a high value, the handover will be triggered too early leading to lower utilization of the WLAN services by the MN. The result shows the benefit of having an adaptive handover threshold to avoid packet loss during handover.
\section{Results for HTS}
\begin{figure*}[h]
\begin{center}
 \includegraphics[scale=0.8]{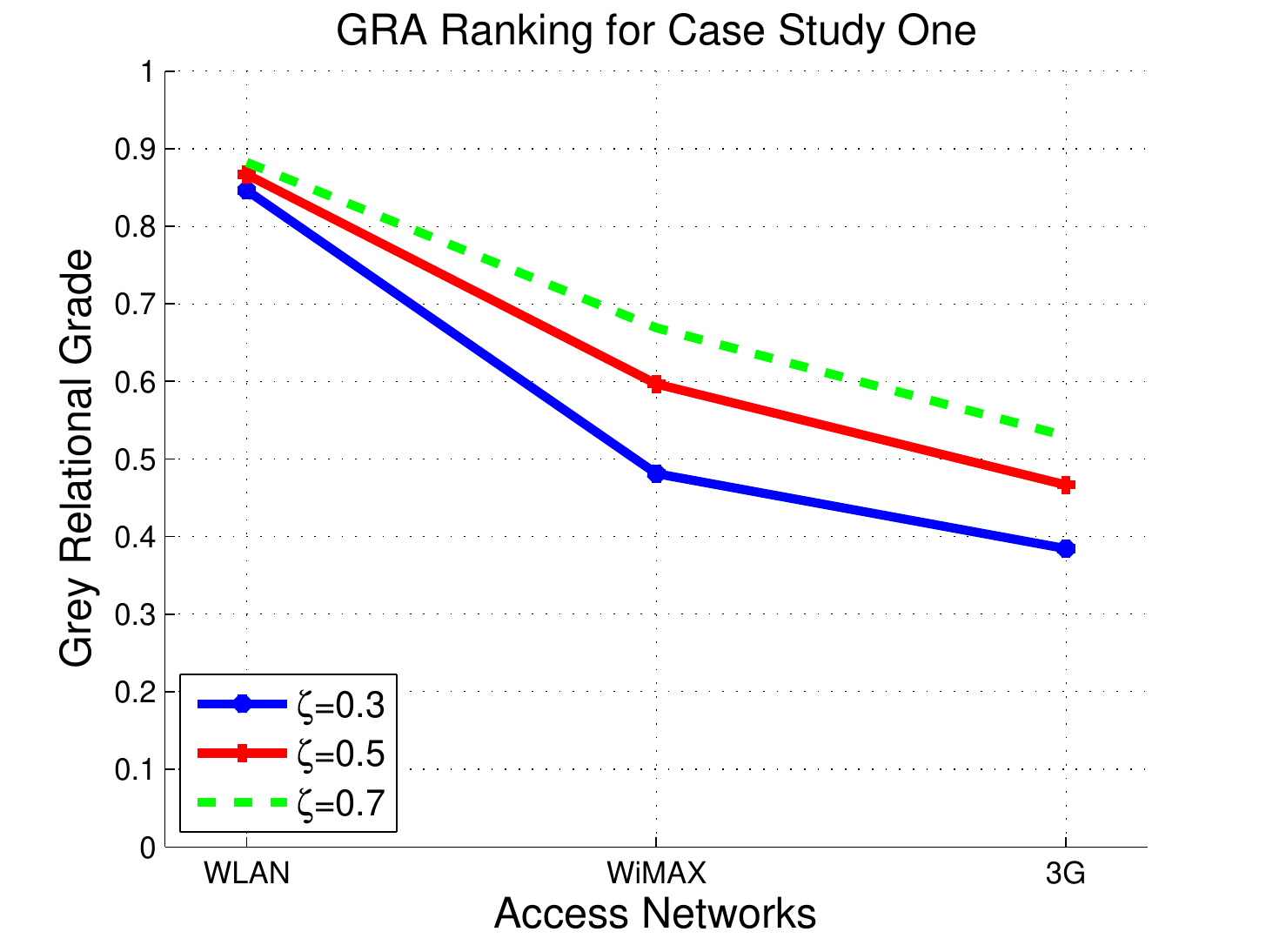}
 \caption{GRA Ranking of Case Study One.}
\label{fig:4e}       
\end{center}
\end{figure*}

\begin{figure*}[h]
\begin{center}
 \includegraphics[scale=0.8]{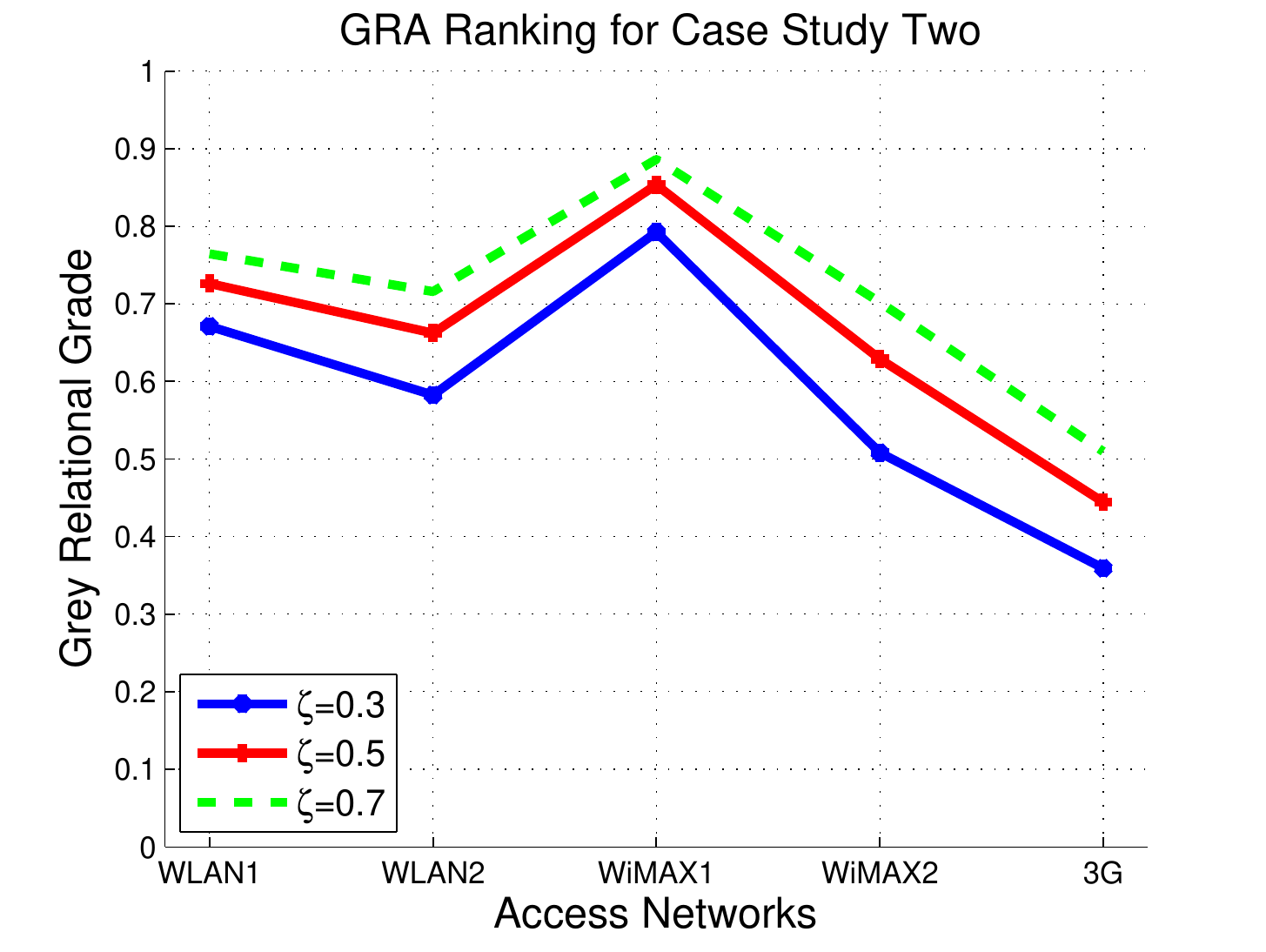}
 \caption{GRA Ranking of Case Study Two.}
\label{fig:4f}       
\end{center}
\end{figure*}

Empirical results show that the GRA approach offers reliable solution when compared with the results obtained from existing methods \cite{RefJ13}. It has the capability of providing differences between access networks which is lacking by most MCDM algorithms \cite{RefJ13}. This study examined the impact of the distinguishing coefficient on the grey relational grade with the coefficient,~$\zeta$ set at values (0.3, 0.5, 0.7) as shown in Fig. \ref{fig:4e} and \ref{fig:4f}, for Case study one and two respectively. The results indicates that the distinguishing coefficient has minimal impact on the GRA rankings, thus, showing the validity of this approach.
From Fig. \ref{fig:4e} (in case study one), the WLAN was ranked first and this implies that when an MN is in the radio range of three access networks, it will be best to perform a handover to the WLAN cell based on the stated performance attributes. On the other hand (in case study two), Fig. \ref{fig:4f} indicates that WiMAX1 access network was ranked best and this makes it the most suitable network for the MN to perform a handover. In both cases, the rankings were constant despite using various distinguishing coefficient. The GRA approach employed in this study provides satisfactory results which will assist system designers in implementing an effective decision making process.

\section{Summary}
In this chapter, performance evaluation was carried out on the proposed vertical handover scheme. Using MATLAB, simulation experiments were done to demonstrate that the proposed handover necessity estimation is able to minimize the number of handover failures and unnecessary handovers. The proposed handover triggering condition estimation approach was able to offer the flexibility of choosing, either for increased WLAN usage or reduced probability of connection breakdown, based on the nature of application used or user preferences. Finally, GRA algorithm was used in the proposed handover target selection approach for optimal network selection.  Combining these three approaches gives a novel scheme which can be used in optimizing vertical handover decisions.

%% file: chap5.tex
\chapter{Conclusion and Future Work}
\label{chp:5}
\newpage
\section{Conclusion}
{\Huge $\mathbb{T}$}his paper has presented new models for realistic renderings of the WLAN coverage area. The proposed geometric-based model for HNE and HTCE combines and extends theoretical results from previous mathematical analysis conducted by several researchers. The resulting model is probabilistic and based on various network parameters which include the random varying cell radius, the traverse angle,~$\theta$, the velocity of the MN. This model is also unique in the sense that it can simulate different coverage scenarios with respect to dwell time of the MN in the WLAN cell. As all parameters of the models were derived from extensive geometric and probability analysis, they correctly simulate the actual behavior of the MN traversing a WLAN coverage area. From results obtained, we arrive at the conclusion that shadow fading has minimal effect on vertical handover models. Our future work will consider the effects of small scale fading due to multi-path.

The models presented were validated by comparing simulated results with works of other researchers under similar conditions. The quality of these simulations qualitatively matched the actual behaviors of MN traversing a realistic WLAN cell. To the best of my knowledge, this is the first geometric-based model that considers an amoebic cell structure for simulating the probability of handover. The model was successful in minimizing Probability of Unnecessary Handover,~$P_u$, Probability of Handover Failure,~$P_f$ and Probability of connection breakdowns,~$P_{Break}$. Using the GRA algorithm, the scheme was able to provide a computationally faster way of making efficient handover decisions that will increase user satisfaction based on multiple performance attributes. The geometric-based models for HNE and HTCE are also the first models of its kind in the existing literatures.


%% file: biblio.tex
\begin{singlespace}
\renewcommand{\bibname}{References}

\bibliographystyle{plain}
\end{singlespace}